\documentclass{aa}  

\usepackage[utf8]{inputenc}

\usepackage{graphicx}
\usepackage{natbib}
\usepackage{caption}

\usepackage{booktabs} 

\usepackage{multicol, blindtext} 

\usepackage{longtable}
\usepackage{lscape}

\graphicspath{{./Pictures_paper_jpg_format/}} % Specifies the directory where pictures are stored

\usepackage[dvipsnames]{xcolor}
\usepackage{lipsum}

\usepackage{txfonts}
\begin{document}

   \title{The CARMENES search for exoplanets around M dwarfs}

   \subtitle{A super-Earth planet orbiting HD 79211 (GJ\,338\,B) \thanks{Tables \ref{CARMENES_rv_measurments}--\ref{tab:ref astrometric data} are only available in electronic form at the CDS via anonymous ftp to {\tt cdsarc.u-strasbg.fr (130.79.128.5)}
or via \tt http://cdsweb.u-strasbg.fr/cgi-bin/qcat?J/A+A/}
   }

   \author{E. Gonz\'alez-\'Alvarez \inst{\ref{inst:CSIC-INTA1}}
        \and M.~R.~Zapatero~Osorio\inst{\ref{inst:CSIC-INTA1}}
        \and  J.~A.~Caballero\inst{\ref{inst:CSIC-INTA1}}
        \and J.~Sanz-Forcada \inst{\ref{inst:CSIC-INTA1}}
        \and V.~J.~S.~B\'ejar \inst{\ref{inst:IAC},\ref{inst:ULL}}
        \and L.~Gonz\'alez-Cuesta \inst{\ref{inst:IAC},\ref{inst:ULL}}
        \and S.~Dreizler \inst{\ref{inst:IAG_goett}} 
        \and F.~F.~Bauer \inst{\ref{inst:IAA-CSIC}} 
        \and E.~Rodr\'iguez \inst{\ref{inst:IAA-CSIC}} 
        \and L.~Tal-Or \inst{\ref{inst:TelAviv},\ref{inst:IAG_goett}}
        \and M.~Zechmeister \inst{\ref{inst:IAG_goett}} 
        \and D.~Montes \inst{\ref{inst:UCM}} 
        \and M.~J.~L\'opez-Gonz\'alez \inst{\ref{inst:IAA-CSIC}} 
        %%------------grupo2----------------------
     	\and I.~Ribas\inst{\ref{inst:CSIC-ICE},\ref{inst:IEEC}}  
    	\and A.~Reiners\inst{\ref{inst:IAG_goett}}
    	\and A.~Quirrenbach\inst{\ref{inst:zah_lsw}}
     	\and P.~J.~Amado\inst{\ref{inst:IAA-CSIC}}
     	%%------------grupo3----------------------
     	\and G.~Anglada-Escud\'e \inst{\ref{inst:IAA-CSIC},\ref{inst:QMUL}}  
     	\and M.~Azzaro \inst{\ref{inst:CAHA}} 
     	\and M.~Cort\'es-Contreras  \inst{\ref{inst:CSIC-INTA1}} 
     	\and A.~P.~Hatzes \inst{\ref{inst:TLS}} 
     	\and T.~Henning \inst{\ref{inst:MPIA}} 
     	\and S.~V.~Jeffers \inst{\ref{inst:IAG_goett}}
     	\and A.~Kaminski \inst{\ref{inst:zah_lsw}} 
     	\and M.~K\"urster \inst{\ref{inst:MPIA}} 
     	\and M.~Lafarga \inst{\ref{inst:CSIC-ICE},\ref{inst:IEEC}}
     	\and J.~C.~Morales \inst{\ref{inst:CSIC-ICE},\ref{inst:IEEC}}
     	\and E.~Pall\'e \inst{\ref{inst:IAC},\ref{inst:ULL}}
     	\and M.~Perger \inst{\ref{inst:CSIC-ICE},\ref{inst:IEEC}} 
     	\and J.~H.~M.~M.~Schmitt \inst{\ref{inst:HS}}
          }
         
			% 1 
             \institute{Centro de Astrobiolog\'ia (CSIC-INTA), Carretera de Ajalvir km 4, 28850 Torrej\'on de Ardoz, Madrid, Spain\label{inst:CSIC-INTA1}
             %2
             \and Instituto de Astrof\'isica de Canarias, Av. V\'ia L\'actea s/n, 38205 La Laguna, Tenerife, Spain\label{inst:IAC}
             %3
             \and Departamento de Astrof\'isica, Universidad de La Laguna, 38206 La Laguna, Tenerife, Spain\label{inst:ULL}
             %4
             \and Institut f\"ur Astrophysik, Georg-August-Universit\"at, Friedrich-Hund-Platz 1, 37077 G\"ottingen, Germany\label{inst:IAG_goett}
              %5
              \and Instituto de Astrof\'isica de Andaluc\'ia (IAA-CSIC), Glorieta de la Astronom\'ia s/n, 18008 Granada, Spain\label{inst:IAA-CSIC}
              %6
				\and Department of Physics, Ariel University, Ariel, 40700, Israel \label{inst:TelAviv}   
              %6.5
              \and Departamento de F\'isica de la Tierra y Astrof\'isica \& UPARCOS-UCM (Ins. de F\'isica de Part\'iculas y del Cosmos de la UCM), Facultad de Ciencias F\'isicas, Universidad Complutense de Madrid, 28040 Madrid, Spain\label{inst:UCM}
              %7
              \and Institut de Ci\`encies de l'Espai (ICE, CSIC), Campus UAB, c/ de Can Magrans s/n, 08193 Bellaterra, Barcelona, Spain\label{inst:CSIC-ICE}
              %8
				\and Institut d’Estudis Espacials de Catalunya (IEEC), 08034 Barcelona, Spain \label{inst:IEEC} 
				%9
			 \and Landessternwarte, Zentrum f\"ur Astronomie der Universt\"at Heidelberg, K\"onigstuhl 12, 69117 Heidelberg, Germany\label{inst:zah_lsw}
              %10
             \and School of Physics and Astronomy, Queen Mary University of London, 327 Mile End Road, London E1 4NS, UK \label{inst:QMUL}
               %11
                \and Centro Astron\'omico Hispano-Alem\'an (CSIC-MPG), Observatorio Astron\'omico de Calar Alto, Sierra de los Filabres, 04550 G\'ergal, Almer\'ia, Spain\label{inst:CAHA}
%                %12
             \and Th\"uringer Landessternwarte Tautenburg, Sternwarte 5, 07778 Tautenburg, Germany\label{inst:TLS}
%                %13
               \and  Max-Planck-Institut f\"ur Astronomie, K\"onigstuhl 17, 69117 Heidelberg, Germany\label{inst:MPIA}
%                 %14
                 \and Hamburger Sternwarte, Gojenbergsweg 112, 21029 Hamburg, Germany\label{inst:HS}  
                    }

   \offprints{Esther Gonz\'alez-\'Alvarez \\ \email{egonzalez@cab.inta-csic.es}
}
   \date{Received 05 November 2019 / Accepted 06 March 2020}

%\abstract{}{}{}{}
% 5 {} token are mandatory

  \abstract
   {}
   {We report on radial velocity time series for two M0.0\,V stars, GJ\,338\,B and GJ\,338\,A, using the CARMENES spectrograph, complemented by ground-telescope photometry from Las Cumbres and Sierra Nevada observatories. We aim to explore the presence of small planets in tight orbits using the spectroscopic radial velocity technique.
}
   {We obtained 159 and 70 radial velocity measurements of GJ\,338\,B and A, respectively, with the CARMENES visible channel between 2016 January and 2018 October. We also compiled additional relative radial velocity measurements from the literature and a collection of astrometric data that cover 200\,a of observations to solve for the binary orbit.}
   {We found dynamical masses of 0.64\,$\pm$\,0.07\,M$_\odot$ for GJ\,338\,B and 0.69\,$\pm$\,0.07\,M$_\odot$ for GJ\,338\,A. The CARMENES radial velocity periodograms show significant peaks at 16.61\,$\pm$\,0.04 d (GJ\,338\,B) and 16.3$^{+3.5}_{-1.3}$\,d (GJ\,338\,A), which have counterparts at the same frequencies in CARMENES activity indicators and photometric light curves. We attribute these to stellar rotation. GJ\,338\,B shows two additional, significant signals at 8.27\,$\pm$\,0.01 and 24.45\,$\pm$\,0.02\,d, with no obvious counterparts in the stellar activity indices. The former is likely the first harmonic of the star's rotation, while we ascribe the latter to the existence of a super-Earth planet with a minimum mass of 10.27$^{+1.47}_{-1.38}$\,$M_{\oplus}$ orbiting GJ\,338\,B. We have not detected signals of likely planetary origin around GJ\,338\,A. 
   }
   {GJ\,338\,Bb lies inside the inner boundary of the habitable zone around its parent star. It is one of the least massive planets ever found around any member of stellar binaries. The masses, spectral types, brightnesses, and even the rotational periods are very similar for both stars, which are likely coeval and formed from the same molecular cloud, yet they differ in the architecture of their planetary systems.
   }


\maketitle

%
%--------------------------------------------------------------------
%--------------------------------------------------------------------
%--------------------------------------------------------------------
%--------------------------------------------------------------------
%--------------------------------------------------------------------
%--------------------------------------------------------------------
%-------------------------------------------------------------------

\section{Introduction}
\label{Introduction}

Despite the significant number of discoveries of planetary systems 
with M dwarf primaries \cite[e.g.,][]{2012ApJ...751L..16A,2013A&A...556A.110B,2015A&A...575A.119A,2017A&A...605A..92S,2019A&A...622A.193A}, the properties and the statistics of planets hosted by low-mass stars remain poorly constrained. We are far from understanding such fundamental questions as how planetary systems form and how their architecture changes with the mass of the central star. 

Various existing radial velocity (RV) surveys focus on M dwarfs for many reasons, including: ($i$) they are the most abundant stellar population in the solar vicinity \citep{2006AJ....132.2360H}, ($ii$) the ocurrence of small planets (typically with sizes 1--4 times that of the Earth) is increasingly higher toward late spectral types at all orbital periods explored by the {\sl Kepler} mission (\citealt{2013ApJ...767...95D, 2014ApJ...791...10M, 2015ApJ...807...45D, 2015ApJ...798..112M, 2016MNRAS.457.2877G}), ($iii$) planets of low-mass stars can be detected easily with the radial velocity technique, and ($iv$) the appealing possibility of spatially resolving terrestrial planets in the habitable zone of the nearest stars using next-generation instrumentation \citep[e.g.,][]{2012ApJS..201...15H,2013ApJ...767...95D,2013EPJWC..4703006S}. 

However, M dwarfs also have their own issues, for example, they have large convective regions and are on average more active than solar-like stars \citep{1997A&A...327.1114L, 2005ApJ...621..398O}. It is crucial to study the stellar activity of M dwarfs together with the analysis of the presence of a planetary signal in their RV time series data. Many works \citep[e.g.,][]{2009ApJ...691..342H,2010MNRAS.408..475H,2015MNRAS.452.2745S,2016ApJ...821L..19N,2019A&A...621A.126D} have discussed how the rotational periods of M dwarfs often coincide with the orbital periods of planets in the expected habitable zone of these stars. The closeness between these periodicities represent an observational challenge since the signal of the planet has to be disentangled from the magnetic activity contribution of the star \citep{2018A&A...615A..69D}.

Instruments such as CARMENES \citep{2016SPIE.9908E..12Q} and GIARPS \citep[GIANO-B+HARPS-N,][]{2017EPJP..132..364C} minimize the stellar activity problem by obtaining simultaneous optical (VIS) and near-infrared (NIR) data. The RV signal of planetary origin is independent of the observed wavelength, whereas the amplitude of the signal due to stellar activity may strongly depend on wavelength \citep[typically larger at shorter wavelengths; e.g.,][]{2006ApJ...644L..75M, 2008A&A...489L...9H,2018A&A...613A..50C}. The main goal of CARMENES is to discover and characterize Earth-like planets around an initial sample of about 300 M dwarfs \citep{2018A&A...612A..49R}. To date, the program has already confirmed eight planet candidates from large-scale photometric and spectroscopic surveys \citep[e.g.,][]{2018A&A...609A.117T,2018AJ....155..257S} and detected more than ten new planets \citep[e.g.,][]{2018A&A...609L...5R, 2018A&A...618A.115K, 2018A&A...620A.171L, 2019A&A...628A..39L, 2018Natur.563..365R,2019A&A...622A.153N,2019A&A...624A.123P,2019A&A...627A..49Z,2019Sci...365.1441M}.

Here, we present the detailed CARMENES RV analysis of the M0.0\,V stars GJ\,338\,A (HD~79210) and GJ\,338\,B (HD~79211), a wide binary system with similar mass stellar components. In Section~\ref{sec:target_stars}, we introduce the binary and the known properties of each stellar member. Section~\ref{sec:Observations} presents all new and literature RV data employed in this paper for the study of the binary orbit and the exploration of the presence of small planets around each star. We also present the recently obtained optical photometry that helped us confirm the rotation period of the stars independently of the spectroscopic measurements. In Section~\ref{sec:Analysis}, we provide the analysis of the stellar binary orbit by combining all available astrometry and literature RVs with the main goal of determining the dynamical masses of the two stars, and the detailed study of the CARMENES RV data and the new photometry aimed at identifying low-mass planets. The properties of the newly discovered planet orbiting GJ\,338\,B are given in Section~\ref{sec:Best-fit parameters}. A brief discussion on the implications of this finding and the conclusions of this paper appear in Sections~\ref{Discussion} and~\ref{Summary and conclusions}.

%--------------------------------------------------------------------
%--------------------------------------------------------------------
%--------------------------------------------------------------------
%--------------------------------------------------------------------
%--------------------------------------------------------------------
%--------------------------------------------------------------------
\section{Target stars}
\label{sec:target_stars}

GJ\,338\,B is a nearby, bright M0.0\,V star at a distance of 6.334\,pc \citep[\textit{Gaia} Data Release 2, DR2;][]{2018A&A...616A...1G}. As described in detail by \cite{1973AJ.....78.1093A} and \cite{1987ApJ...317..343M}, this star, together with the twin M0.0\,V star HD\,79210 (GJ\,338\,A), forms the high common proper motion pair $\Sigma$ Struve STF~1321 (WDS~J09144+5214, ADS 7251).
Occasionally, they have been classified as K7\,V stars, but here we follow the latest classification by \cite{2015A&A...577A.128A}. These authors determined spectral type M0.0\,V for both stars (actually, GJ\,338\,A is the M0.0\,V spectral standard star used in that work). The most updated stellar parameters of GJ\,338\,B and~A are compiled in Table~\ref{stellar_parameters_1}.

The projected separation between GJ~338\,A and~B is 108.54\,au (17\farcs2). \cite{2017A&A...597A..47C} did not find any close companions to either GJ~338\,A or~B during their high-resolution  imaging survey. The Washington Double Star catalog \citep{2001AJ....122.3466M} tabulates two additional components C and D in the system STF~1321, located to date at slightly over 2\,arcmin to the northeast and southeast, respectively, of the central, much brighter pair AB. With the latest {\em Gaia} DR2 data we confirm that C and D are background stars located further away and with very different proper motions as compared to our system.\footnote{''C'': \object{2MASS~J09143546+5242095}; ''D'': \object{TYC~3806--1033--1}.}

According to \cite{1955MNRAS.115..187V}, the first orbital parameter determinations of the pair STF~1321 were done by \cite{Hopmann54} and \cite{Guntzel-Lingner54}. % (today Commission G1). 
The former author estimated a long orbital period, $P_{\rm orb}$ = 1555\,a, and a large semimajor axis, $a$ = 22\farcs36, whereas the latter author published a short period, $P_{\rm orb}$ = 687\,a and a small semimajor axis $a$ = 16\farcs56.
Since these first determinations, the orbital parameters of the visual binary have been revisited on a few occasions \citep{1980AZh....57.1227K,1994AZh....71..278R,1996ARep...40..795K,2012A&A...546A..69M} but none has surpassed the quality of the fit found by \cite{1972AJ.....77..759C}, who measured an intermediate period, $P_{\rm orb}$ = 975\,a and semimajor axis $a$ = 16\farcs72 (ORB6 orbit grade = 4, following \citealt{2001AJ....122.3472H}). \cite{1972AJ.....77..759C} also derived stellar masses of 0.41$\pm$0.03\,M$_\odot$ and 0.73$\pm$0.05\,M$_\odot$ for GJ~338~A and~B, respectively. These mass derivations, where the secondary is significantly more massive than the primary, is reversed with respect to what is expected from the brightness of stars: GJ\,338\,B, supposedly the most massive component, is $\sim$0.1 mag fainter than GJ\,338\,A in all optical and near-infrared filters. These results are also in apparent disagreement with the latest, more similar mass determinations by \citealt{2019A&A...625A..68S} (see below). In particular, these authors determined masses of 0.58--0.61\,M$_\odot$ for each stellar component using different methods (see also \citealt{2012ApJ...757..112B}, \citealt{2014MNRAS.443.2561G}, and \citealt{2018A&A...612A..49R}). 

The chemical composition analysis available in the literature reveals that GJ~338~A and~B have a slightly sub-solar metallicity with reported measurements ranging from $\rm[Fe/H]=0.0$ to $-0.35$\,dex \citep{2009ApJ...704..975J,2014ApJ...791...54G,2013A&A...551A..36N,2016A&A...585A..64S}. The recent study by \cite{2018A&A...615A...6P} reported a metallicity of 0.07$\pm$0.16 and $-$0.03$\pm$0.16 for GJ~338~A and~B, which lies close to the solar metallicity but also agrees with previous metallicity derivations within the quoted error bars. Based on these results, no significant difference in the chemical composition of GJ~338~A and~B is expected.

Both stars were cataloged as probable kinematically young stars in the Local Association by \cite{2011MNRAS.410..190T}. This is consistent with their membership in the Galactic young disk (age $\le$ 1\,Ga) reported by \cite{miriam_thesis}, who pointed out that, despite their Galactic kinematics indicating youth, other photometric and spectroscopic observables of GJ~338~A and~B do not support a young age. \citet{2013A&A...555A..11E} reported the non-detection of mid-infrared flux excesses in GJ\,338\,B using {\sl Herschel} data, thus discarding the presence of a dusty disk, which is compatible with the most likely stellar age of a few Ga.

GJ\,338\,A and GJ\,338\,B are active and show intense chromospheric fluxes \citep{2017MNRAS.472.4563M}. This is consistent with the slightly fast rotation of components A and B, which have a projected rotational velocity $v$\,sin\,$i = 2.9\pm1.2$ and $2.3\pm1.5$\,km\,s$^{-1}$ \citep{2005ESASP.560..571G,2018A&A...612A..49R}. GJ\,338\,A and B have been observed in X-rays by \textit{Chandra} on 29/12/2012, and we derived their luminosities at     
$L_{\rm X} =5.2\pm0.3\times10^{27}$ and $5.0\pm 0.1\times10^{27}$\,erg\,s$^{-1}$, respectively. From their X-ray luminosities, both components have a similar activity level with values typically found among active stars of their spectral type.

This X-ray luminosity is compatible with an age of 1 to 7\,Ga following the relation by \cite{2011A&A...532A...6S}. Based on the activity index $\log R'_{\rm HK}$ $\sim$\,--4.4 for GJ\,338\,B, \cite{2010A&A...520A..79M} and \cite{2010ApJ...725..875I} estimated that this star is active with an expected radial velocity jitter of about 4--10\,m\,s$^{-1}$.

\begin{table}[t]
\centering
\small
\caption{Stellar parameters of GJ\,338 binary system.}
\label{stellar_parameters_1}
\setlength{\tabcolsep}{1pt}
\begin{tabular}{l c c l}
\hline
\hline
\noalign{\smallskip}
Parameters  & GJ\,338\,A & GJ\,338\,B & Ref.$^{a}$\\
\noalign{\smallskip}	
\hline	
\noalign{\smallskip}

Other name					&		HD 79210	&	HD 79211 & \\
Karmn								&		J09143+526 	&	J09144+526 & AF15a \\
$\rm \alpha$ (J2000) &		09:14:22.78 &   09:14:24.68	& \textit{Gaia DR2}\\ % a) GAIA DR2
$\rm \delta$ (J2000) &		+52:41:11.8	&   +52:41:10.9 & \textit{Gaia DR2}\\
$G$ (mag) &		$6.9689\pm0.0005$	& $7.0477\pm0.0004$ & \textit{Gaia DR2}\\
$J$ (mag)  &		$4.89\pm0.04$	& $4.779\pm0.174$ &  2MASS\\
Spectral type				&		$\rm M0.0\,V$	&	$\rm M0.0\,V$	 & AF15a\\ 

$d$ (pc)	  &	$6.334\pm0.002$		& $6.334\pm0.002$ & \textit{Gaia DR2}\\
$\mu _{\alpha} \cos \delta$ ($\rm mas\,a^{-1}$) &		$-1546.10\pm0.06$ 	& $-1573.12\pm0.06$ & \textit{Gaia DR2}\\ 
$\mu _{\delta}$ ($\rm mas\,a^{-1}$)&		$-569.13\pm0.06$	& $-660.12\pm0.06$ & \textit{Gaia DR2}\\
$U$ ($\rm km\,s^{-1}$)	&		$-42.20\pm0.36$	& $-44.01\pm0.36$ & CC16\\
$V$ ($\rm km\,s^{-1}$)	&		$-14.99\pm0.10$	& $-17.44\pm0.10$ & CC16\\
$W$ ($\rm km\,s^{-1}$)	&		$-23.73\pm0.34$ 	& $-23.10\pm0.34$ & CC16\\
%$V_{\rm r}$ ($\rm km\,s^{-1}$)					& $12.263\pm0.110$	& Rein17\\

$T_{\rm eff}$ (K)	&		$4024\pm51$	&	$4005\pm51$ & Schw19\\
$\log g$ (cgs) 	&	$4.68\pm0.07$		& 	$4.68\pm0.07$ & Schw19\\
{[Fe/H]} (dex) 	&		$-0.05\pm0.16$	&	$-0.03\pm0.16$ & Schw19\\

$L$ ($L_{\odot}$) &		$0.0789\pm0.0038$	& 	$0.0792\pm0.0031$ & Schw19\\ 
$M$ ($M_{\odot}$)     &	$0.69\pm0.07$		& $0.64\pm0.07$  & This work\\  
									 	&	$0.591\pm0.047$		& 	$0.596\pm0.042$ & Schw19\\ 

$R$ ($R_{\odot}$) 	&		$0.58\pm0.02$	& $0.58\pm0.03$ & Schw19\\

$ v \sin i$ ($\rm km\,s^{-1}$) &	$2.9\pm1.2$		& $2.3\pm1.5$ & GG05, Rein18\\
$P_{\rm rot}$ (d) 	&		$16.3^{+3.5}_{-1.3}$	& $16.61 \pm 0.04$ & This work\\
$L_{\rm X}$ ($10^{27}$ erg\,$s^{-1}$) &	$5.2\pm0.3$ 	& $5.0\pm 0.1$ & This work\\

$\log R'_{\rm HK}$	&		...  	& $-4.4$ & MA10\\
Age (Ga)										&		1--7			& 1--7 & This work\\

\hline
\end{tabular}
\tablefoot{$^{a}$ \textit{Gaia} DR2: \cite{2018A&A...616A...1G}; 2MASS: \cite{2006AJ....131.1163S}; AF15a: \cite{2015A&A...577A.128A}; CC16: \cite{miriam_thesis}; GG05: \cite{2005ESASP.560..571G}; Rein18: \cite{2018A&A...612A..49R}; Schw19: \cite{2019A&A...625A..68S}; MA10: \cite{2010A&A...520A..79M}. }
\end{table}

%--------------------------------------------------------------------
%--------------------------------------------------------------------
%--------------------------------------------------------------------
%--------------------------------------------------------------------
%--------------------------------------------------------------------
%--------------------------------------------------------------------
%--------------------------------------------------------------------

\section{Observations}
\label{sec:Observations}

\subsection{Radial velocities from the literature}
\label{sec:Radial velocity from the literature}

Absolute radial velocities of each star are available in a number of publications and archives: \cite{1973AJ.....78.1093A}, \cite{1986ApJS...62..147B}, \cite{1986PASP...98..772B}, and the \textit{Gaia} DR2 catalog \citep{2018A&A...616A...7S}. With the only exception of the \textit{Gaia} DR2 data, all other absolute velocities are affected by rather large error bars (typically $\ge$1 km\,s$^{-1}$) and dispersion, and we did not consider them in our analysis. More recent radial velocities with smaller associated uncertainties (typically from several to tens of m\,s$^{-1}$) come from the ELODIE \citep{1996A&AS..119..373B} and SOPHIE \citep{2008SPIE.7014E..0JP} high-resolution spectrographs. These data were taken as part of the ground-based efforts intended for the calibration of the velocities of the \textit{Gaia} DR2 catalog \citep{2018A&A...616A...7S}. The ELODIE and SOPHIE radial velocities were obtained using different masks at the pipeline level, which had an impact on the determination of the velocity zero points. Unfortunately, the SOPHIE data of GJ\,338\,A and~B were analyzed using different masks per star, thus preventing us from a direct comparison between the two stellar components of the binary. As pointed out by \cite{2018A&A...616A...7S}, the error associated with the determination of the velocity zero point offsets due to the different masks employed during the velocity computations was large for the M dwarfs ($\approx \pm0.4$ km\,s$^{-1}$), and it was not precisely quantified. The ELODIE radial velocities, however, were reduced with the same mask for the two stars, and we used these measurements in our study of the orbital parameters of the stellar pair. Eleven ELODIE radial velocities covered the time interval between January 1995 and February 2000. They have a dispersion of 10--30 m\,s$^{-1}$, which is about one order of magnitude worse than the CARMENES data (Section~\ref{sec:CARMENES radial velocity time series}). Therefore, we do not use these data to explore the presence of small-mass planets around GJ\,338\,A and B. However, these measurements have sufficient quality for obtaining the orbital solution fit of the stellar binary.

\begin{figure*}[!t]
\centering
\begin{minipage}{0.32\linewidth}
\includegraphics[angle=0,scale=0.26]{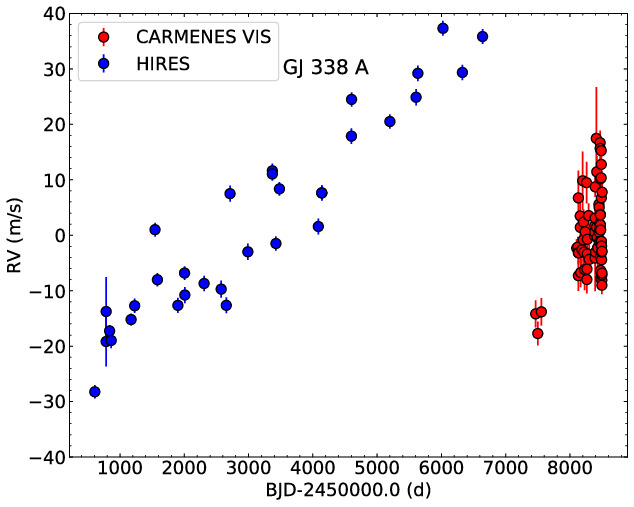}
\end{minipage}
\begin{minipage}{0.32\linewidth}
\includegraphics[angle=0,scale=0.26]{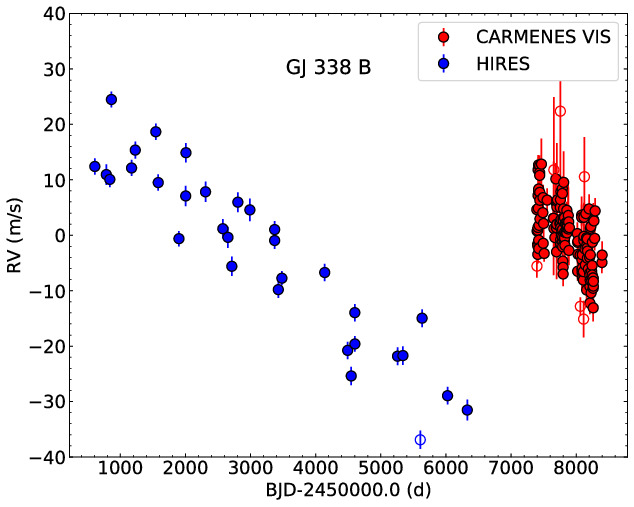}
\end{minipage}
\begin{minipage}{0.32\linewidth}
\includegraphics[angle=0,scale=0.26]{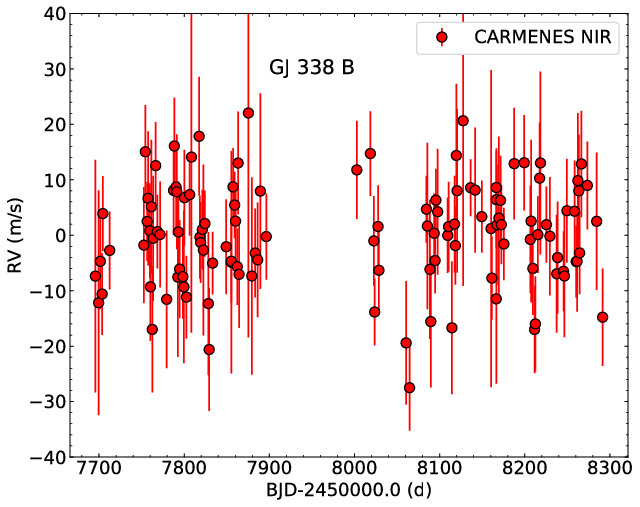}
\end{minipage}
\caption{{\sl Left panel:} GJ\,338\,A HIRES (blue) and CARMENES (red) relative RV measurements at visible wavelengths. {\sl Middle panel:} GJ\,338\,B HIRES (blue) and CARMENES (red) relative RV measurements at visible wavelengths. The time series data of each instrument has been normalized to a mean null velocity. Open circles indicate measurements that deviate more than 2$\sigma$ from the observed velocity trend. {\sl Right panel:} CARMENES relative RVs of GJ\,338\,B in the near-infrared.
}
\label{fig:gj338b_vrad}
\end{figure*}

In addition to the ELODIE data, and for the characterization of the stellar binary orbit, we employed the more than 30 relative radial velocities obtained for each member of the stellar pair with the High Resolution (HIRES) spectrograph \citep{1994SPIE.2198..362V} of the Keck telescope published by \citet{2017AJ....153..208B}. These velocities span a period of 15.7\,a and were later corrected for the presence of systematic effects affecting the nightly zero points (a discontinuous jump caused by interventions on the instrument in 2004 and a long-term drift; see \citealt{2019MNRAS.484L...8T}). The HIRES RVs have associated a mean error bar value of 1.6\,m\,s$^{-1}$. The relative HIRES RVs of GJ\,338\,A and~B given by \cite{2019MNRAS.484L...8T} are displayed in Fig.~\ref{fig:gj338b_vrad}, where the trend due to the orbital motion of the pair around the center of mass of the system is clearly seen for each star. To bring the HIRES relative RVs given by \cite{2019MNRAS.484L...8T} to an absolute RV calibration, we used the corresponding values reported by \citet{2002ApJS..141..503N}, which fully overlap with the first years of HIRES observations. The ELODIE and the HIRES data altogether cover over 20\,a of observations. ELODIE absolute velocities were taken to the HIRES reference by considering the measurements of the two datasets at common observing epochs.

The radial velocity survey of \citet{2016ApJ...822...40G} using the CSHELL spectrograph \citep{1990SPIE.1235..131T} at the NASA Infrared Telescope Facility (IRTF) included observations of GJ\,338\,A and~B at near-infrared wavelengths ($K$ band) between 2014 March and 2014 December. They do not overlap with any of the ELODIE, HIRES, or CARMENES data. \citet{2016ApJ...822...40G} concluded that hot, warm and cool Jupiter planets more massive than 2.3, 4.2, and 20\,M$_{\rm Jup}$ and with orbital periodicities of 1--10, 10--100, and 100--1000\,d, respectively, can be excluded around the B component (our main CARMENES target) with a confidence of 95\%. The CSHELL/IRTF relative velocities are not considered in our planetary search of Section~\ref{sec:Analysis} because they have associated uncertainties of the order of  6--18 m\,s$^{-1}$, that is, more than three times larger than the typical CARMENES error bar. In addition, they are not included in our analysis of the stellar binary because no absolute velocities were reported by the authors and no calibration to absolute data is possible due to the lack of overlapping dates with any other data in hand. 

%--------------------------------------------------------------------
%--------------------------------------------------------------------
%--------------------------------------------------------------------
%--------------------------------------------------------------------
%--------------------------------------------------------------------
%--------------------------------------------------------------------
%--------------------------------------------------------------------

\subsection{CARMENES radial velocity time series}
\label{sec:CARMENES radial velocity time series}

We obtained new RVs for each member of the stellar binary using the CARMENES fibre-fed, \'echelle spectrograph. CARMENES is installed at the 3.5 m telescope of the Calar Alto Observatory in Almer\'ia (Spain). It was specifically designed to deliver high-resolution spectra at optical (resolving power $R = 94,600$) and near-infrared ($R = 80,500$) wavelengths covering from 520 to 1710 nm without any significant gap, using two different spectrograph arms: the VIS and NIR channels, respectively \citep{2016SPIE.9908E..12Q}.

The CARMENES M-dwarf survey started in January 2016. The original sample of stars contained 324 M dwarfs \citep{2016csss.confE.148C, 2018A&A...609L...5R,2018A&A...612A..49R}, including GJ\,338\,B. Raw data are automatically reduced with the CARACAL pipeline \citep{2016SPIE.9910E..0EC}. Relative radial velocities are extracted separately for the VIS and NIR channels using the SERVAL software \citep{2018A&A...609A..12Z}, which computes the average of all individually derived RVs order by order of the ech\'elle observations. The results of the CARMENES planetary survey around M dwarfs reveal that the instrument is able to deliver time series of relative RVs with root mean square ($rms$) values of 3.5 and 9.0\,m\,s$^{-1}$ for the CARMENES VIS and NIR channels \citep{2018A&A...612A..49R}.

The systematic RV errors in spectral orders highly contaminated by telluric contribution, which is the case of CARMENES NIR data, can be increased by unmasked detector defects \citep{2018A&A...612A..49R}. Therefore, for the NIR channel, we carefully selected the orders that were included in the computation of the NIR RVs. These orders, selected to provide the smallest rms of the NIR RVs, were chosen after analyzing the entire CARMENES M-dwarf sample.

%--------------------------------------------------------------------
%--------------------------------------------------------------------
%--------------------------------------------------------------------
%--------------------------------------------------------------------
%--------------------------------------------------------------------
%--------------------------------------------------------------------

\subsubsection{GJ 338 B}
GJ\,338\,B is the main target of this paper. We used 159 CARMENES VIS and 120 NIR relative RVs obtained between 2016 January and 2018 October. There are fewer NIR RVs because we selected only the NIR RVs taken after a technical intervention in the NIR channel in October 2016. GJ\,338\,B was observed with a typical cadence of one RV measurement every few to several days to sample short periodicities. All measurements employed in this work are provided in Tables~\ref{CARMENES_rv_measurments} (VIS) and~\ref{CARMENES_rv_measurments_nir} (NIR). All data are corrected from barycentric motion, secular acceleration, instrumental drift, and nightly zero points \citep{2018A&A...609A.117T}. CARMENES VIS relative RVs of GJ\,338\,B as given in the CARMENES catalog of observations are depicted as a function of observing epoch in the middle panel of Fig.~\ref{fig:gj338b_vrad}. The decreasing trend due to the orbital motion of the binary agrees with the trend delineated by the HIRES relative RVs. The right panel of Fig.~\ref{fig:gj338b_vrad} shows CARMENES NIR relative RVs of GJ\,338\,B. The slope of the NIR data is not as clear as that of the VIS RVs likely due to the larger uncertainties (and therefore, dispersion) of the NIR RVs.

%--------------------------------------------------------------------
%--------------------------------------------------------------------
%--------------------------------------------------------------------
%--------------------------------------------------------------------
%--------------------------------------------------------------------
%--------------------------------------------------------------------

\subsubsection{GJ 338 A}

The primary component of the stellar binary was also observed as part of the CARMENES M-dwarf program with a total of 70 RV measurements in the VIS channel taken between 2016 March and 2019 January. These VIS RVs, also relative and corrected from barycentric motion, secular acceleration, instrumental drift, and nightly zero points, are provided in Table~\ref{Tab:CARMENES_rv_measurements_Acomp}. GJ\,338\,A was not observed regularly at the beginning of the CARMENES M-dwarf project: the first three RVs were obtained more than one year before the remaining RV time series data. After 2017 December 18, GJ\,338\,A was monitored with a similar observing cadence as GJ\,338\,B. CARMENES VIS relative RVs are depicted as a function of observing epoch in the left panel of Fig.~\ref{fig:gj338b_vrad}; they are shown together with the HIRES RV data of \cite{2019MNRAS.484L...8T}. Because the time coverage of the CARMENES data of GJ\,338\,A with a good time sampling is significantly shorter than that of GJ\,338\,B, we could not explore the same orbits as those of our study of GJ\,338\,B. These RVs were also used to study the orbital motion of the stellar pair (see Section~\ref{Orbital_solution}) and to act as a reference (for its similarity in spectral type) in our search for low-mass planets around GJ\,338\,B.

%--------------------------------------------------------------------
%--------------------------------------------------------------------
%--------------------------------------------------------------------
%--------------------------------------------------------------------
%--------------------------------------------------------------------
%--------------------------------------------------------------------

\subsection{Photometric time series}
\label{phot}

\begin{table}[]
\caption{Photometric seasons available for GJ\,338\,A and B.}
\label{tab:phot:propert}
\centering  

\setlength{\tabcolsep}{1pt}
\begin{tabular}{l c c c c c }
\hline
\hline
\noalign{\smallskip}
Obs.$^a$  & Filter & Season & $\Delta$T & $\rm N_{obs}$ & $\sigma$\\
 		&		& & (d) & & (mmag) \\
\noalign{\smallskip}
\hline
\noalign{\smallskip}

LCO & $V$ & 2018 Mar 5--27 & 22 & 1830 & 14.4 \\
LCO & $i$  & 2018 Mar 5--27 & 22 & 2100 & 13.4  \\
LCO & $V$  & 2018 Dec 2 -- 2019 Feb 27 &     87  & 350 &  10.7\\
LCO & $B$    & 2018 Dec 2 -- 2019 Feb 27 &   87   & 330 &  8.0\\
SNO & $V$   & 2019 Feb 22 -- 2019 Apr 13 &   50   & 819 & 6.1 \\
SNO & $B$   & 2019 Feb 22 -- 2019 Apr 13&   50  & 839 &  5.9\\

\hline
\end{tabular}
\tablefoot{$^{a}$ Observatories -- LCO: Las Cumbres; SNO: Sierra Nevada.}
\end{table}

\begin{figure}[]
\includegraphics[width=\columnwidth]{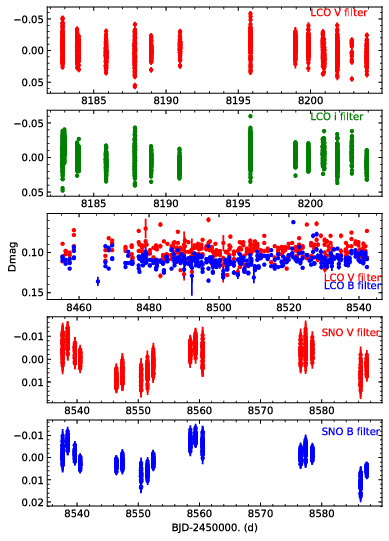}
\caption{Differential photometry (GJ\,338\,B -- GJ\,338\,A) light curves obtained with the LCO and SNO telescopes. First and second panels correspond to the first photometric observing season covering a time interval of 22 days; the third panel illustrates the light curve for the second observing season covering 87 days of continuous monitoring; and the last observing season spanning 50 days is shown in the fourth and fifth panels. With the exception of the third panel, all photometry is artificially set to a mean value of zero magnitude. 
}
\label{fig:all_phot_data}
\end{figure}

We obtained optical photometric time series of both GJ\,338\,A and B contemporaneous to the CARMENES data, aiming to derived the stellar rotational periods via the differential photometry technique. Images were acquired with telescopes of Las Cumbres Observatory Global Telescope \citep[LCO,][]{2013PASP..125.1031B} and Sierra Nevada Observatory (SNO). The 40-cm LCO telescopes used in our program are equipped with a 3k$\times$2k SBIG CCD camera with a pixel scale of 0$\farcs$571 providing a field of view of 29$\farcm$2$\times$19$\farcm$5. The 90-cm Ritchey-Chrétien SNO telescope has a VersArray 2k$\times$2k CCD camera with a square field of view of 13$\farcm$2$\times$13$\farcm$2 \citep[pixel scale of 0$\farcs$39 on sky,][]{2010MNRAS.408.2149R}. We applied a 2$\times$2 binning to the detector. Raw frames were reduced following standard procedures at optical wavelengths, that is, bias subtraction and flat-field correction. Bad pixels were conveniently masked using  well-tested masks provided by the observatories. In the case of LCO images, we used the {\tt banzai} reduction products \citep{2018SPIE10707E..0KM}.

There were two photometric campaigns using LCO telescopes between 2018 March and 2019 February, and a third campaign using the SNO telescope from 2019 February to April. All three are summarized in Table~\ref{tab:phot:propert}, where we provide observing filters, the number of days covered per observing season, the number of photometric measurements, and the dispersion of the differential photometry. In the first photometric season, from 2018 March 5 to 27 (22 days of continuous monitoring), we acquired a total of 1800 and 2300 usable images of GJ\,338\,A and~B using the $V$ and $i$ filters and a  2\,$\times$\,2 binning of the detector. The on-source integration time was set at 1 s, and 100 consecutive images were typically acquired per filter and per night. On a few nights, the 100-frame sequence was acquired twice or three times. Weather conditions were variable from night to night as was also the seeing, with values varying between 1\farcs0 and 5\farcs0. The root-mean-square ($rms$) of the data are 14.4 mmag for the $V$-band light curve and 13.4 mmag for the $i$-band time series. The two light curves artificially normalized to zero magnitude are depicted in the top panels of Fig.~\ref{fig:all_phot_data}.

In the second LCO campaign, from 2018 December through 2019 February (almost three months of continuous monitoring), we followed a slightly different observing strategy: between three and five images were typically obtained in the $B$ and $V$ filters throughout the observing nights resulting in more than 350 and 330 usable $B$- and $V$-band images whose differential photometry is shown in the middle panel of Fig.~\ref{fig:all_phot_data} (no normalization to zero magnitude was applied here). In this second season, the $rms$ of the $V$-band light curve (10.7 mmag) is smaller than that of the first season. The third photometric season (2019 February 22 through April 13, that is, 50 days of nearly continuous monitoring) was acquired using the $B$ and $V$ filters immediately after the last LCO campaign. Typically, a total of 50 images were acquired with the SNO telescope per night. Exposure time changed between 6 s and 20 s, depending on the seeing conditions. The SNO light curves are illustrated in the bottom panels of Fig.~\ref{fig:all_phot_data}, and have an $rms$ of 6.1\,mmag ($V$) and 5.9\,mmag ($B$). Also the LCO $B$-band light curve shows a smaller dispersion than the $V$-band data.

We used a circular apertures with radius of seven and 20 pixels to determine the LCO and SNO photometry of GJ\,338\,A and~B, respectively. We checked that these apertures minimized the $rms$ of the resulting light curves. The field of view registered by the LCO and SNO detectors is unfortunately devoid of bright stars that can act as reference sources for the differential photometry. Therefore, we derived the differential magnitude between the two components of the stellar binary (component B -- component A). Because both stars share very similar spectral type, we did not apply any color correction to the differential photometry. With such a procedure, it is not possible to discern which star is the source of any photometric variability, or if the two stars are variable simultaneously. 

Given the large number of contemporaneous $B$, $V$, and $i$ individual measurements available for both stellar components, we also derived differential $B-V$ and $V-i$ color light curves and found that GJ\,338\,A and~B indeed show different colors, and that these colors appear rather stable throughout all monitoring campaigns. The relative color differences, although small, are significant in terms of the achieved photometric precision. GJ\,338\,B has $B-V$ and $V-i$ indices that are, on average,  12.5\,$\pm$\,0.6 and 34.2\,$\pm$\,0.7\,mmag redder than those of GJ\,338\,A, where the error bars were obtained as the quadratically-summed photometric dispersions of both bands divided by the squared root of the number of measurements. These differences can be interpreted as both stars having slightly different temperatures (GJ\,338\,B would be cooler than GJ\,338\,A). Another possibility is a differing chemical composition of the stellar atmospheres (higher metallicity induces redder colors), but we give less credit to this scenario because the two stars are likely coeval and were probably formed from the same molecular cloud. GJ\,338\,B is also slightly fainter than the A component ($\Delta B = $ 110.2 $\pm$ 0.4, $\Delta V = $ 97.7 $\pm$ 0.6, and $\Delta i = $ 65.6 $\pm$ 0.2 mmag), which combined with its redder nature may indicate that both stars have similar but not identical masses. The photometric properties of GJ\,338\,B (fainter and redder) suggest that this star is slightly less massive than GJ\,338\,A; this is compatible with the results presented in Section~\ref{Orbital_solution} but strongly disagrees with the mass derivations of \cite{1972AJ.....77..759C}.

GJ\,338\,A and~B were also photometrically observed by the All-Sky Automated Survey for Supernovae survey \citep[ASAS-SN,][]{2017PASP..129j4502K}. The ASAS-SN light curves show strong peak to peak variations. This behavior is far from what is seen in the LCO and SNO data. The ASAS-SN photometry was automatically derived from a 2-pixel radius aperture (i.e., about two FWHM in diameter) and the background was estimated from a 7--10\,pixel radius annulus surrounding each of the two stars. The pixel scale of ASAS-SN detector is 8\farcs0. Given the projected spatial separation of the stellar binary ($\rho \sim 17\farcs2$), it becomes obvious that the ASAS-SN parameters for extracting the photometry of the system were not adequate and yield contaminated and useless light curves for GJ\,338\,A and~B.

%--------------------------------------------------------------------
%--------------------------------------------------------------------
%--------------------------------------------------------------------
%--------------------------------------------------------------------
%--------------------------------------------------------------------
%--------------------------------------------------------------------

\section{Analysis}
\label{sec:Analysis}

\subsection{Stellar binary orbital solution}
\label{Orbital_solution}

\begin{figure}[]
\includegraphics[width=\columnwidth]{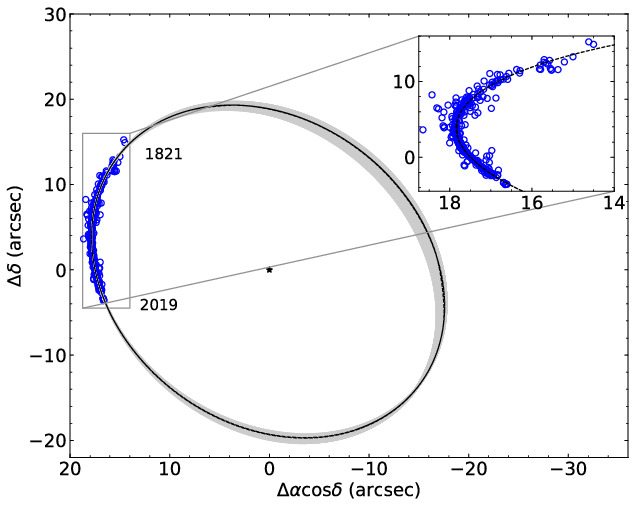}
\includegraphics[width=1.03\columnwidth]{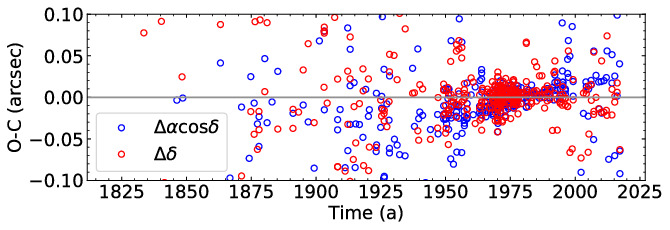}
\caption{{\sl Top panel:} observed astrometric positions of GJ\,338\,B (blue dots) around the star GJ\,338\,A (black star located at 0,0 coordinates) expressed in $\Delta\delta$ vs. $\Delta\alpha \cos \delta$. The black ellipse (dashed line) corresponds to the adopted Keplerian solution, which lies close to the region of the largest concentration of possible MCMC solutions. The oldest data and most recent astrometric data correspond to the year 1821 and 2019, respectively. The inner panel is a zoom-in around the astrometric data. Only the adopted orbital solution is shown with the black dashed line. {\sl Bottom panel:} residuals ($\pm$0\farcs11 in $\Delta\alpha \cos \delta$ and $\pm$0\farcs14 in $\Delta\delta$) after removing the adopted orbital solution plotted as a function of observing epoch ($\Delta\alpha \cos \delta$: blue dots; $\Delta\delta$: red dots).
}
\label{fig:gj338b_astrometry}
\end{figure}

In an attempt to constrain the orbital parameters of the stellar pair GJ\,338\,A and~B, we considered all astrometric and spectroscopic data available to us in the pre-CARMENES era. As for the astrometry, we used the angular separations and position angles compiled by the Washington Double Star Catalog \footnote{\url{https://www.usno.navy.mil/USNO/astrometry/optical-IR-prod/wds/WDS}} \citep[WDS,][]{2001AJ....122.3466M}. This dataset provides astrometric measurements from mid 1821 through the first quarter of 2019. We completed the astrometric data using the SNO images from Section~\ref{phot}. By employing the 2MASS \citep{2006AJ....131.1163S} coordinates of all sources detected in the SNO frames, with the exception of the high proper motion targets GJ\,338\,A and~B, we obtained the astrometric solution for seven different epochs in 2019, thus extending the time coverage of the Washington Double Star Catalog by two additional years. We found a plate scale of 0.388094 $\pm$ 0.000030\,arcsec/pix (the pixels have the same length in the $x$- and $y$-axis within the quoted uncertainty) and that the SNO frames are 1.5 deg off the standard north--east orientation. This was corrected from the final measurements. We derived that the angular position of GJ\,338\,B with respect to the A component has changed by about 58\,deg in 198\,a of available astrometric observations. All astrometric data, including our latest derivations, in the form of angular separation ($\rho$) and position angle ($\theta$) are given in Table\ref{tab:astrometric data} and displayed in Fig.~\ref{fig:gj338b_astrometry}. As for the RVs, we used the ELODIE and HIRES absolute RVs of GJ\,338\,A and~B as explained in Section~\ref{sec:Radial velocity from the literature}.

To find the best Keplerian solution to the orbit of the stellar pair, and based on our knowledge of the system, any orbital solution has to reproduce the observed astrometric and radial velocity data and to be compliant with the following: a) The total mass of the system should fall in the interval 1.03--1.38 M$_{\odot}$, which is obtained from the 1$\sigma$ smallest and largest individual values assigned to each component in Table~B.1 of \citet{2019A&A...625A..68S}. According to these authors, the most likely total mass of the system is $1.187 \pm 0.063$ M$_{\odot}$. b) The two stellar components have very similar spectral type, effective temperature, magnitude, and luminosity \citep{2018A&A...615A...6P, 2019A&A...627A.161P}. Therefore, we expect the mass ratio, $q$, to be close to unity. In fact, \citet{2019A&A...625A..68S} determined averaged masses of $0.591 \pm 0.047$ M$_{\odot}$ for GJ\,338\,A and $0.596 \pm 0.042$ M$_{\odot}$ for GJ\,338\,B (errors are also averaged from those given by the authors). These determinations were based on the analysis of CARMENES spectra and do not rely on astrometric or radial velocity observations.

\begin{figure}[!ht]
\includegraphics[width=\columnwidth]{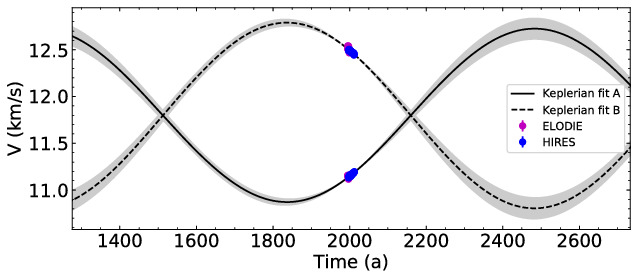}
\includegraphics[width=\columnwidth]{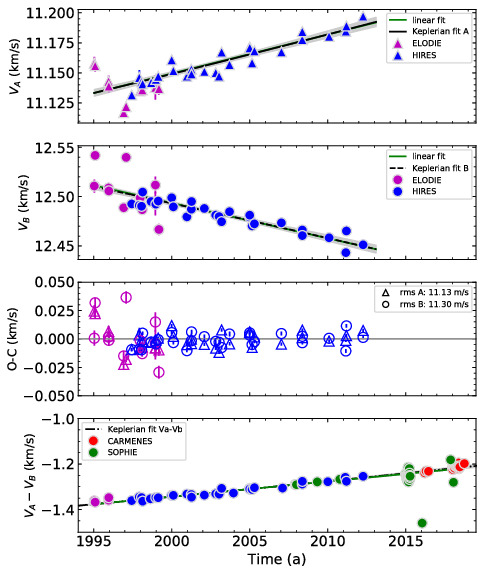}
\caption{{\sl First panel:} radial velocity data and the joint astrometric and spectroscopic solution (black line) of the stellar binary system formed by GJ\,338\,A and~GJ\,338\,B. The uncertainties are depicted with a gray color. ELODIE and HIRES absolute RVs are also shown. {\sl Second and third panels:} zoom-in around the RVs of GJ\,338\,A and~GJ\,338\,B, respectively. The straight line fit to the data is plotted as a green line. Given the high level of agreement between the adopted Keplerian solution and the straight line, the latter remains nearly hidden by the former in the two panels. {\sl Fourth panel:} residuals of the RV observations minus the adopted Keplerian solution are shown with different symbols for the two stars. {\sl Bottom panel:} ELODIE, HIRES, SOPHIE and CARMENES differential radial velocity curve ($v_A-v_B$) as a function of observing epoch. The values of the adopted Keplerian solution, which were fit using the ELODIE and HIRES data only, are shown with a black dashed line, errors in gray, and the straight line fit to the data is depicted with a green line. All four panels, from the second to the bottom one, have the same scale in the x-axis.
}
\label{fig:gj338b_RV_abs}
\end{figure}

The orbital parameters of the stellar pair GJ\,338\,A and~GJ\,338\,B were obtained by simultaneously fitting the astrometric and RV data. Our code is based on {\tt PyMC3}, which is a flexible and high-performance model building language and inference engine that scales well to problems with a large number of parameters, and on the {\tt exoplanet} toolkit \cite[e.g.,][]{exoplanet:theano}, which extends {\tt PyMC3}'s language to support many of the custom functions and distributions required when fitting our datasets \citep{exoplanet:pymc3}. {\tt Exoplanet} is also designed to provide the building blocks for fitting many datasets using the Markov Chain Monte Carlo (MCMC) simulation approach. We set to 100,000 the number of MCMC simulations to derive the parameter distribution for the orbit of the binary; we allowed MCMC to explore values of orbital parameters in all possible ranges (e.g., eccentricity from 0.0 to almost 1.0), except for the orbital period and the amplitude of the RV curves, where we explored values in the intervals 800--5000\,a and 0.4--5.0\,km\,s$^{-1}$. The posterior distributions of the fitted parameters are graphically shown in the corner plot of Fig.~\ref{fig:my_corner_output.png}, where it becomes apparent that the distribution of the solutions is narrow with clear peaks for most of the orbital parameters. The eccentricity distribution shows a peak at 0.0 with a long tail towards slightly larger values, yet our solution strongly indicates that the orbit is nearly circular with an eccentricity likely smaller than $e = 0.1$. More degenerated is the determination of the orbital inclination angle. This is not a surprising result because the available astrometric data cover less than 20\%~of the actual orbit of the stellar binary, which obviously is insufficient to find a tight solution.

Here, we adopted the orbital parameters mean values of the MCMC posterior distributions as the ``best'' Keplerian solution describing the observed astrometry and absolute RVs altogether. All adopted orbital parameters are given in Table~\ref{tab:gj338_orbital_param}. The mean values are very close to the peaks of the distributions shown in the corner plot of Fig.~\ref{fig:my_corner_output.png}. We found that the binary has a likely orbital period of 1295\,$\pm$\,184\,a, which is significantly larger than that of \cite{1972AJ.....77..759C} but closer to the determination by \cite{Hopmann54}, and a semi-major axis of 130.9\,$\pm$\,5.1\,au (or 20\farcs66\,$\pm$\,0\farcs80). The error bars were determined by differenciating the various mean solutions obtained when removing data points from the long astrometric dataset. Figure \ref{fig:gj338b_astrometry} shows the observed astrometric positions of GJ\,338\,B around GJ\,338\,A between 1821 and 2019, the adopted Keplerian ellipse, and the residuals after removing the adopted orbital solution from the data. The residuals have a dispersion of $\pm$0\farcs11 and $\pm$0\farcs14 in right ascension ($\alpha$) and declination ($\delta$), respectively, which are close to the uncertainties of the best quality astrometry in our dataset. Furthermore, the residuals are flat over time, which indicates that the solution is acceptable for the covered time interval. With this solution, a total mass of 1.33\,$\pm$\,0.17\,M$_\odot$ is obtained for the stellar pair, which is consistent at the 1$\sigma$ level with the most likely value obtained by \citet{2019A&A...625A..68S}. The indivial mass of each binary component was then determined from the total mass and amplitude of the RV curves (mass ratio, $q = K_A/K_B = M_B/M_A = 0.93 \pm 0.13$). We found dynamical masses of 0.69\,$\pm$\,0.07\,M$_\odot$ for GJ\,338\,A and 0.64\,$\pm$\,0.07\,M$_\odot$ for GJ\,338\,B. 

Figure \ref{fig:gj338b_RV_abs} shows the ELODIE and HIRES absolute RVs and the best Keplerian fit. The top panel illustrates 1.5 full orbital periods; the absolute RV data considered in this work cover a tiny fraction ($<$10\,\%) of the actual orbital period. The second and third panels of Fig.~\ref{fig:gj338b_RV_abs} show an enlargement of the previous panel centered on the observed velocities of GJ\,338\,A and B. Both the adopted Keplerian solution and a straight line fit to the RV data are included in the two panels. Given the small time coverage of the data, there is no perceptible difference between the Keplerian fit and the straight line (at the level of less than 1\,m\,s$^{-1}$, that is, about one order of magnitude smaller than the RV uncertainties). The fourth panel of Fig.~\ref{fig:gj338b_RV_abs} illustrates the velocity residuals (observations minus the fit), which have a dispersion of 11.13\,m\,s$^{-1}$ (GJ\,338\,A) and 11.30\,m\,s$^{-1}$ (GJ\,338\,B).

As a consistency check of the quality of our adopted Keplerian fit, we produced the bottom panel of Fig.~\ref{fig:gj338b_RV_abs} that displays the differential velocity ($v_A - v_B$) of the binary members as a function of time using ELODIE, HIRES, SOPHIE, and CARMENES data, thus expanding the time coverage of the RV dataset by an additional 5.7\,a for a total of 23.7\,a. We assumed that the RV measurements of GJ\,338\,A and B are coeval if they were acquired on the same night or within a week time. With this condition, we identified 29 HIRES, 5 ELODIE, 39 SOPHIE, and 23 CARMENES RV pairs for GJ\,338\,A and B. This plot is in principle independent (or at least it minimizes the systematics) of the absolute RVs provided that all RVs are obtained using the same mask. For this purpose, we re-reduced all publicly available ELODIE, SOPHIE and CARMENES spectra using the SERVAL software \citep{2018A&A...609A..12Z} and the same template of GJ\,338\,B for the two stars, thus removing the velocity zero point problem mentioned in Section~\ref{sec:Radial velocity from the literature}. The adopted Keplerian solution nicely reproduces the differential velocity curve of the ELODIE and HIRES RVs and also extrapolates equally well towards the more recent SOPHIE and CARMENES data.

Another consistency check of the validity of the adopted Keplerian solution is given by the determination of the binary mass ratio through the measurement of the slopes of the combined ELODIE and HIRES RV curves of GJ\,338\,A and~B. The mass ratio obtained in this way is $q=0.91 \pm 0.11$, which is compatible at the 1$\sigma$ level with the value derived from the ratio of the RV amplitudes ($K_A, K_B$) of the adopted Keplerian fit.

\begin{figure}[]
\includegraphics[width=\columnwidth]{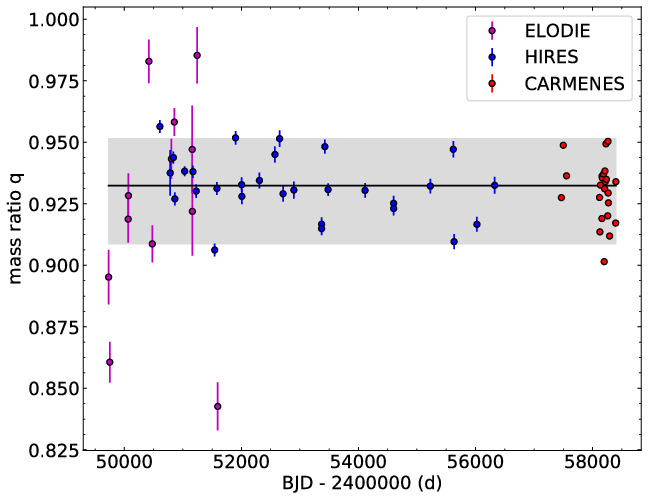}
\caption{Determination of the stellar mass ratio, $q=M_B/M_A$, using ELODIE, HIRES, and CARMENES VIS differential radial velocities (see Section~\ref{Orbital_solution}). The horizontal, solid line corresponds to $q=0.929\pm0.019$, where the uncertainty gray area accounts for the observed dispersion of the $q$ values.}
\label{fig:mass_ratio_q}
\end{figure}

Another method to derive the mass ratio of the binary is based on the equation $q=(v_A - \gamma) / (\gamma - v_B)$ \citep{1941ApJ....93...29W}, where $\gamma$ stands for the systemic radial velocity. Because the mass ratio has to be a constant, we forced a flat (zero slope) $q$ versus time as shown in Figure~\ref{fig:mass_ratio_q}. We employed the ELODIE and HIRES absolute RVs previously used in the Keplerian solution determination and enlarged the data coverage adding the CARMENES relative RVs calibrated to absolute value using the adopted Keplerian solution. The derived values are as follows: $q = 0.930 \pm 0.021$, $\gamma = 11.797 \pm 0.001$\,km\,s$^{-1}$. The small uncertainties associated with $q$ and $\gamma$ reflect the precision of the method, which leads to the conclusion that GJ\,338\,A is likely the most massive component of the stellar binary, as expected from its higher luminosity and bluer optical colors. This result is opposed to that of \cite{1972AJ.....77..759C}. By using the astrometric total mass of the stellar system, we obtained individual masses of 0.69 $\pm$ 0.03\,M$_\odot$ (GJ\,338\,A) and 0.64 $\pm$ 0.03\,M$_\odot$ (GJ\,338\,B), which are basically identical to the results from the first approach (combined astrometric and spectroscopic fit) and are consistent with the mass determinations by \cite{2019A&A...625A..68S} at the 1$\sigma$ level, but slightly larger by 8.3--9.4\,\%~than those reported by these authors. 

From now on, we use the stellar mass determinations from the combined astrometric and spectroscopic fit. Table~\ref{tab:gj338_orbital_param} summarizes the derived orbital parameters, including the individual mass determinations and the mass ratio determinations from the spectroscopic data only.

\begin{table*}[]
\centering
\caption{Orbital parameters for the GJ\,338 stellar binary system.}
\label{tab:gj338_orbital_param}
\setlength{\tabcolsep}{5pt}
\begin{tabular}{l c c c}
\noalign{\smallskip}
\hline
\hline
\noalign{\smallskip}
\multicolumn{4}{c}{Combined astrometric and spectroscopic solution} \\ 
\noalign{\smallskip}
\hline
\noalign{\smallskip}
Parameter  & Description (Unit) & Adopted solution \\
\noalign{\smallskip}	
\hline	
\noalign{\smallskip}

$\alpha$\dotfill & Semi-major axis (arcsec)\dotfill & 20.66 $\pm$ 0.80 \\
$a$\dotfill	& Semi-major axis (au)\dotfill & 130.9  $\pm$ 5.1 \\
$T_0$\dotfill	& Time of periastron passage  (a)\dotfill & 1513 $\pm$ 61 \\
$e$\dotfill & Eccentricity\dotfill & 0.01$^{+0.15}_{-0.01}$\\
$i$\dotfill  & Inclination (deg)\dotfill  & 37 $\pm$ 12 \\
$\Omega$\dotfill & Argument of ascending node (deg)\dotfill & 32.4 $\pm$ 1.8 \\
$\omega$\dotfill & Argument of periastron (deg)\dotfill& 109.1 $\pm$ 6.0 \\
$P$\dotfill & Period (a) \dotfill & 1295 $\pm$ 180 \\
$K_A$ \dotfill & Amplitude (km/s) \dotfill & 0.93 $\pm$ 0.12 \\
$K_B$ \dotfill & Amplitude (km/s) \dotfill & 0.99 $\pm$ 0.12 \\
$M$ \dotfill  & Total mass ($\rm M_{\odot}$)\dotfill & 1.33 $\pm$ 0.17\\
$q$ \dotfill & Mass ratio, $K_A/K_B$ \dotfill & $0.93 \pm 0.10$ \\
$M_A$ \dotfill & Mass of the primary ($\rm M_{\odot}$) \dotfill & $0.69 \pm 0.07$ \\
$M_B$ \dotfill & Mass of the secondary ($\rm M_{\odot}$) \dotfill & $0.64 \pm 0.07$\\
$\gamma$ \dotfill & Systemic velocity (km\,s$^{-1}$) \dotfill & $11.798 \pm 0.001$ \\

\noalign{\smallskip}	
\hline	\hline	
%\end{tabular}
%\begin{tabular}{l c c }
\noalign{\smallskip}	
\multicolumn{4}{c}{Mass ratio$^{(a)}$ from equation $q=(v_A - \gamma) / (\gamma - v_B)$}\\ 
\noalign{\smallskip}
\hline	
\noalign{\smallskip}

$q$ \dotfill & Mass ratio, $M_B/M_A$ \dotfill & $0.930 \pm 0.021$ &\\
$\gamma$ \dotfill & Systemic velocity (km\,s$^{-1}$) \dotfill & $11.797 \pm 0.001$\\

\noalign{\smallskip}	
\hline	
\end{tabular}
\tablefoot{\centering $^{(a)}$ Based on the HIRES absolute RV calibration.}
\end{table*}

%--------------------------------------------------------------------
%--------------------------------------------------------------------
%--------------------------------------------------------------------
%--------------------------------------------------------------------
%--------------------------------------------------------------------
%--------------------------------------------------------------------

\subsection{RV detrending}
\label{detrending}

Before investigating the presence of planets around each member of the GJ\,338 stellar system, the Keplerian orbital solution (Table\ref{tab:gj338_orbital_param}) was subtracted from the observed HIRES and CARMENES optical RVs. As a consistency check, we determined the negative (and positive) slope independently for the HIRES and CARMENES datasets, finding very similar results at the 1$\sigma$ level. We concluded that given the short time coverage of the RV data and the long orbital period of the binary, the subtraction of a straight line instead of the Keplerian curve would not change the results reported next.

Our second step was to remove RV outliers from the GJ\,338\,B datasets by applying a 2$\sigma$-clipping algorithm, resulting in 1 HIRES RV and 6 CARMENES VIS RVs rejected from the lists of 32 and 159 independent measurements, respectively. Outliers typically have large error bars (due to, e.g., short exposures) and are marked in Fig.~\ref{fig:gj338b_vrad} and all other RV figures of this paper. The ``flattened'' and 2$\sigma$-clipping corrected CARMENES VIS RV dataset of GJ\,338\,B has a standard deviation of 4.2\,m\,s$^{-1}$, which is slightly smaller than the dispersion of the HIRES velocities (5.4\,m\,s$^{-1}$). The mean error bar value of the individual CARMENES VIS measurements provided by SERVAL is 2.4\,m\,s$^{-1}$, that is, almost a factor of two lower than the standard deviation. The origin of this RV excess variations is explored in the following section. 

We also removed the orbital trend from the CARMENES NIR data of GJ\,338\,B and applied the 2$\sigma$-clipping for eliminating RV outliers after which we find a standard deviation of 9.2\,m\,s$^{-1}$. This value coincides with the typical dispersion of CARMENES NIR RVs of other M-type stars. Because of the larger $rms$ of the NIR data, this channel does not deliver more convincing results than the VIS channel. Consequently, the NIR RVs were not used except when necessary. 

The same procedure was applied to the CARMENES dataset of GJ\,338\,A resulting in the 2-$\sigma$-clipping removal of 4 VIS RVs out of a total of 70 measurements. The detrended  RV curve of GJ\,338\,A has a standard deviation of 5.5\,m\,s$^{-1}$.

Besides RV precision, the cadence of the observations is also a critical ingredient for performing a proper analysis of temporal variability. In this regard, the CARMENES data were taken every few to several days while the target was visible from the Calar Alto Observatory. This observing frequency contrasts with the cadence of the HIRES observations (3--4 RV measurements per year). As a result, the CARMENES scheduled observations are ideal for exploring planets in short- to intermediate-period orbits \citep{2017A&A...604A..87G}, while the HIRES data are useful for identifying planets at longer orbital periods. Mixing the two datasets might not produce better results given the different data quality and temporal cadence. In what follows, we analyze the CARMENES VIS data independently and employ the HIRES and CARMENES NIR RVs to check the results.

\begin{figure}[]
\includegraphics[width=\columnwidth]{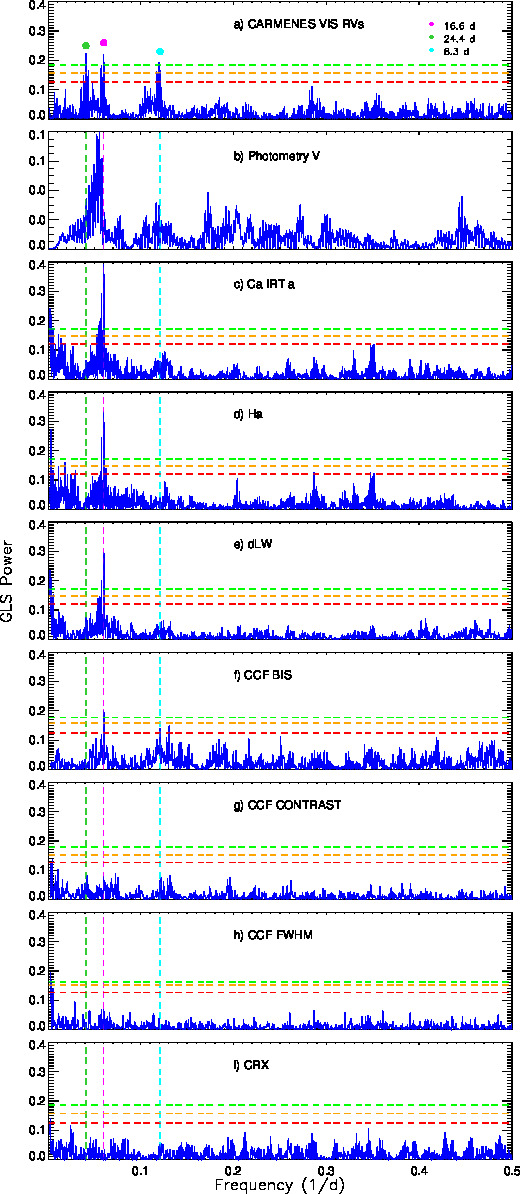}
\caption{GLS periodograms for GJ\,338\,B activity and RV data. In all panels,  the horizontal dashed lines indicate FAP levels of 10\%~(red), 1\%~(orange), and 0.1\%~(green). {\sl Panel (a):} CARMENES RVs from the VIS channel. The three highest peaks at 8.3, 16.6, and 24.4\,d are labeled and their positions are marked in all other panels with vertical dashed lines.  {\sl Panel (b):} $V$-band photometry. {\sl Panels (c--e):} chromospheric line indices of the infrared Ca\,{\sc ii} triplet, H$\alpha$, and dLW. {\sl Panels (f--h):} FWHM, contrast, and bisector velocity span from the CCF analysis. {\sl Panel (i):} chromatic RV index (CRX). 
}
\label{fig:gj338b_activity}
\end{figure}

%--------------------------------------------------------------------
%--------------------------------------------------------------------
%--------------------------------------------------------------------
%--------------------------------------------------------------------
%--------------------------------------------------------------------
%--------------------------------------------------------------------
%--------------------------------------------------------------------

\begin{figure*}[!h]
\centering
\begin{minipage}{0.45\linewidth}
\includegraphics[angle=0,scale=0.55]{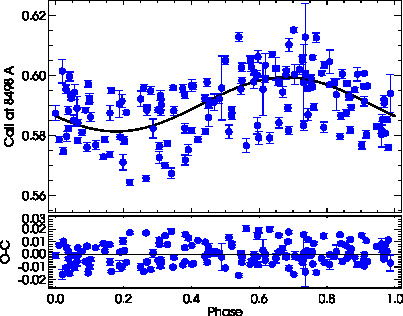}
\end{minipage}
\begin{minipage}{0.45\linewidth}
\includegraphics[angle=0,scale=0.55]{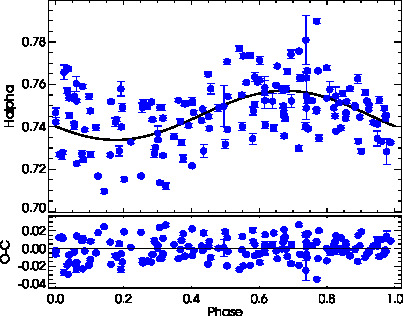}
\end{minipage}
\caption{GJ~338~B. Phase-folded time series of activity indicators  Ca\,{\sc ii} $\lambda$8498\,\AA~({\em left}) and H${\alpha}$ $\lambda$6563\,\AA~({\em right}) and residuals. The solid curve depicts the best sinusoidal fit that modulates the rotation period at 16.6\,d. 
}
\label{fig:gj338b_activity_phase}
\end{figure*}

\subsection{Stellar activity}
\label{sec:Activity}

The profile of the stellar lines can be changed by photospheric and chromospheric activity, which has an impact on accurate RV measurements. M-type dwarfs  typically have intense magnetic fields  \citep{2012AJ....143...93R,2014ApJ...794..144R,2016ApJ...821L..19N,2017ApJ...834...85N,2019A&A...626A..86S}, and it is expected that the related magnetic activity leaves a detectable trace in the high resolution spectra. Various groups have focused on the identification of spectral indicators of stellar activity \citep{2000A&AS..146..103M,2011A&A...534A..30G, 2012A&A...541A...9G,2018A&A...614A.122T} with the main goal of disentangling activity-induced RV variations and planetary signals. Two commonly-used chromospheric activity indicators are based on measurements of the H$\alpha$ $\lambda$6562.82\,\AA~and Ca\,{\sc ii} H and K $\lambda \lambda$3933.7, 3968.5\,\AA~lines. The blue Ca\,{\sc ii} H and K lines are not covered by the CARMENES data; instead, the Ca\,{\sc ii} infrared triplet $\lambda \lambda \lambda$8498, 8542, 8662\,\AA~(IRT) is fully covered by the VIS channel. This triplet is also well known for its sensitivity to stellar activity as reported by \cite{2011MNRAS.414.2629M}. 

The CARMENES SERVAL pipeline provides measurements for a number of spectral features that can reveal stellar activity, including H$\alpha$, Ca\,{\sc ii} IRT, the differential line width (dLW), and the chromatic index (CRX), as defined by \citet{2018A&A...609A..12Z}. The latter measures the RV--$\log \lambda$ correlation, and it is used as an indicator of the presence of stellar active regions. In addition, the pipeline computes the cross-correlation function (CCF) for each CARMENES spectrum using a weighted binary mask that was built from coadded observations of the star itself. As explained by \cite{2018A&A...609L...5R}, the CCFs are fitted with a Gaussian function to determine the contrast, the full width at half maximum (FWHM), and the bisector span (BIS), all of which are useful for studying stellar activity-related properties.

\paragraph{GJ\,338\,B.} All CARMENES spectroscopic activity indicators of GJ\,338\,B are given in Table~\ref{CARMENES_rv_measurments}. Their associated Generalized Lomb-Scargle periodograms \citep[GLS;][]{2009A&A...496..577Z} are depicted in various panels of Fig.~\ref{fig:gj338b_activity}, where we also include the false alarm probability (FAP) levels of 10, 1, and 0.1\,\%~from 10,000 bootstrap randomizations of the input data. Peaks of the GLS periodograms with FAP level $<$0.1\%~are considered significant. The GLS method of analysis is particularly indicated for unevenly time series. With the exception of the CCF contrast, FWHM, and CRX indices, there is one single significant peak in all other activity indicators centered at $\sim$0.060\,d$^{-1}$ ($\sim$16.6\,d). The Ca\,{\sc ii} IRT and H$\alpha$ indices folded in phase with this period are shown in Fig.~\ref{fig:gj338b_activity_phase}: all of them have a sinusoid-like variation (dLW and BIS indices also show similar patterns).

We used the GLS peak at 16.6\,d as the initial period for finding the best sinusoidal curve that minimizes the residuals of the activity indices presented in Fig.~\ref{fig:gj338b_activity_phase}. We found that the best period is 16.61 $\pm$ 0.04\,d, after averaging all individual determinations. We also checked that the variations of the activity indices do not correlate with the RVs. The shape of these variations and the fact that CRX and FWHM indices do not show the 16.6-d peak may be related to the origin and nature of the stellar variability that is inducing this signal in the CARMENES VIS RVs (see below and \citealt{2019A&A...623A..44S}). H${\alpha}$ and Ca\,{\sc ii} IRT are mainly chromospheric indicators while CRX is related to the photosphere. We ascribed the 16.6-d period to the chromospheric rotation cycle of GJ\,338\,B. This period determination differs from the values tabulated by \cite{2011ApJ...743...48W}, who measured a photometric period of 10.17\,d. However, later \cite{2019A&A...623A..24F} reported a tentative period of 16.6$\pm$0.5\,d and 17.4$\pm$1.0\,d using CARMENES H$\alpha$ and Ca\,{\sc ii} IRT activity indices, respectively. These values coincide with the 16.61$\pm$0.04\,d determined here and we agree with these authors that CARMENES Ca\,{\sc ii} IRT and H$\alpha$ activity indices are well suited for stellar period searches.

As a side note, \cite{1973AJ.....78.1093A} measured the Ca\,{\sc ii} emission-line velocities of GJ\,338\,B and~A and found them variable. They discussed that these variations were caused by the intrinsic multiple nature of each star, and determined a ``binary'' period of $\sim$16.47\,d for GJ\,338\,B (the one with the highest significance and amplitude of $\sim$5 km\,s$^{-1}$ according to the authors). Later on, \cite{1987ApJ...317..343M} obtained more data that did not confirm the proposed periods and did not suggest any variation in velocity at all, thus rejecting the notion that GJ\,338\,B is a binary star itself. The CARMENES RVs support this latter result. However, it is possible that \cite{1973AJ.....78.1093A} detected the rotation cycle of GJ\,338\,B because they were using emission lines very sensitive to stellar activity. 

\begin{figure}[]
\includegraphics[width=\columnwidth]{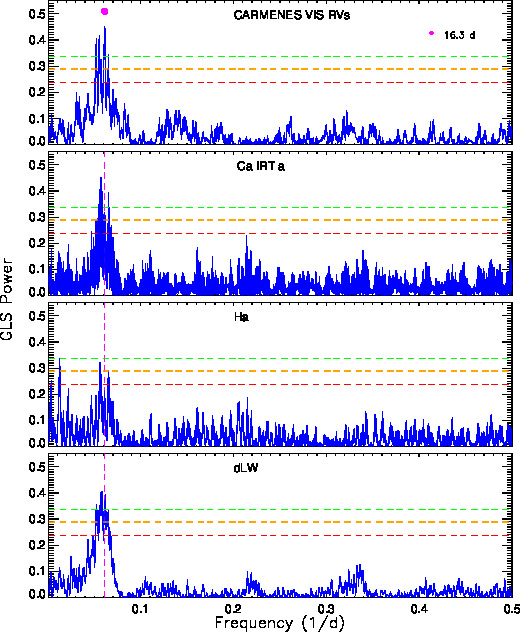}
\caption{GLS periodograms for GJ 338 A. In all panels,  the horizontal dashed lines indicate FAP levels of 10\%~(red), 1\%~(orange), and 0.1\%~(green). {\sl Top panel:} CARMENES VIS RVs. {\sl Second through fourth panels:} CARMENES Ca\,{\sc ii}  $\lambda$8498\,\AA, H$\alpha$, and dLW spectroscopic activity indicators. All activity indicators and the RV data show a significant, broad peak at around 16.3$^{+3.5}_{-1.3}$\,d. This is likely associated with the rotation period of GJ\,338\,A.}
\label{fig:gj338a_multiplot}
\end{figure}

\paragraph{GJ\,338\,A.} We analyzed the CARMENES VIS RVs of GJ\,338\,A in a similar manner as for the B component and found that the GLS periodograms of the H$\alpha$, dLW, and Ca\,{\sc ii} IRT activity indices (see Fig.~\ref{fig:gj338a_multiplot}) show a significant peak at 16.3$^{+3.5}_{-1.3}$\,d (above FAP level $<$0.1\,\%). All RV and activity indices measurements are given in Table~\ref{Tab:CARMENES_rv_measurements_Acomp}. We verified that there is no confusion in the identification of the two stellar sources in the CARMENES catalog. The first evidence is that GJ\,338\,A RV data show a trend that is opposite to that of GJ\,338\,B, thus indicating that we are dealing with the two stellar components of the binary. The second piece of evidence is that the absolute RVs differ between the two stars. We attributed the periodicity of 16.3\,d to the rotational cycle of GJ\,338\,A. Indeed, this rotation period is surprisingly close to that of GJ\,338\,B. It might hint that both M0-type stars, which were likely born from the same molecular cloud and are consequently coeval, have followed a very similar angular momentum evolution.

\paragraph{Photometry.}  For the present analysis, we studied the light curves separately per filter and per observing campaign and all seasons altogether ($V$-band data only, which is the only filter in common to the three campaigns). The differential light curve was obtained by dividing the fluxes of GJ\,338\,B and GJ\,338\,A because there is no other bright star in the field that can act as a reference source (see Section~\ref{phot}). The GLS periodogram of the entire $V$-band light curve is shown in the second panel (from the top) of Fig.~\ref{fig:gj338b_activity}, where a peak at around 18\,d is detectable. The highest peaks in the GLS periodograms of the $B$- and $i$-band light curves lie between 16 and 20\,d, but they are not significant. Also apparent in these periodograms is the peak at 8--10\,d, which is likely due to the first harmonic of the primary signal. With the exception of the $V$-band data, we concluded that no obvious, significant peak can be extracted from the periodograms of the LCO and SNO light curves. From the analysis of the separated seasons of the $V$-filter data, we found a characteristic frequency in the interval 15.8--26.9\,d, which covers the likely rotation periods of the two stars. The lack of an intense peak can be explained by photometric stability (or stellar variability that is below our detectability limit), by anti-correlated stellar variability of similar amplitude (thus producing a net amplitude close to zero in the differential light curve), or by the combination of two stellar variabilities of similar low amplitude and close, but not equal, periodicities, which could attenuate the true individual periodicities. The first hypothesis is the least likely one because there is stellar activity from the CARMENES spectroscopic data. With the current photometry, we could not discern the true scenario and therefore, we were not able to derive individual photometric periodicities for GJ\,338\,B and GJ\,338\,A.

%GLS RV all in one figure
\begin{figure}[!ht]
\includegraphics[width=\columnwidth]{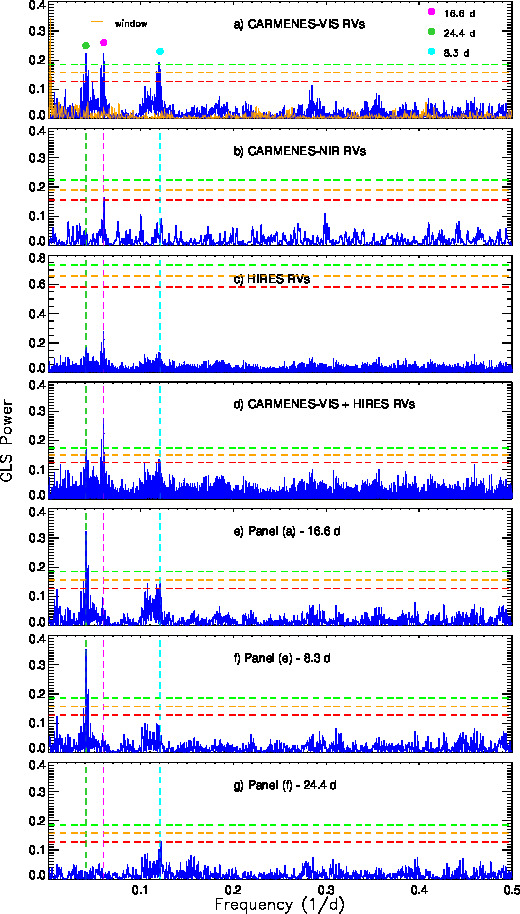}
\caption{GLS periodograms of the RV data of GJ\,338\,B. {\sl Panel (a):} periodogram of the CARMENES VIS RVs is plotted as a blue, solid line and the window function of the RV data is shown with an orange line. The three strongest signals are marked in this panel and in all  other panels. The horizontal dashed lines indicate FAP levels of 10\,\% (red), 1\,\% (orange), and 0.1\,\% (green). {\sl Panel (b):} peridogram of the CARMENES NIR RVs. {\sl Panel (c):} periodogram of the HIRES RV measurements. {\sl Panel (d):} periodogram of the combined HIRES $+$ CARMENES VIS RV data. {\sl Panel (e):} periodogram of the CARMENES VIS RV residuals after removing the best-fit that modulates the stellar activity at $P_{\rm rot}=16.6$\,d. {\sl Panel (f):} periodogram of the CARMENES VIS RV residuals after removing the stellar rotation and its first harmonic at 8.3\,d. {\sl Panel (g):} periodogram of the CARMENES VIS RV residuals after removing all three significant signals.
}
\label{fig:gj338b_GLS_carmenes-eps-converted-to.jpg}
\end{figure}

%--------------------------------------------------------------------
%--------------------------------------------------------------------
%--------------------------------------------------------------------
%--------------------------------------------------------------------
%--------------------------------------------------------------------
%--------------------------------------------------------------------
\begin{figure}[]
  \includegraphics[width=\columnwidth]{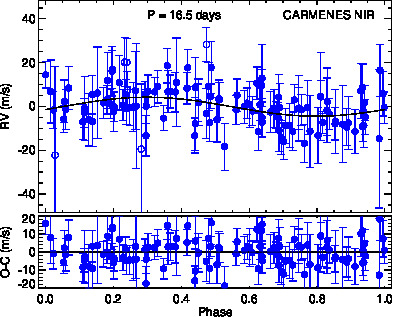}
     \caption{Phase-folded CARMENES NIR RVs (blue dots) to 16.54$\pm$0.06\,d. {\sl Top panel:} the black line depicts the best-fit that modulate the activity. Open circles stand for those points that deviate more than 2$\sigma$. The amplitude is 4.3$\pm$1.7\,m/s. {\sl Bottom panel:} residuals' $rms$ is 7.5\,m/s. }
     \label{Fig_NIR_phase}
\end{figure}

\subsection{Frequency content of the radial velocities of GJ\,338\,B }
\label{Radial velocity analysis}

The GLS periodogram of the CARMENES VIS RVs of GJ\,338\,B is illustrated in the top panels of Figs.~\ref{fig:gj338b_activity} and~\ref{fig:gj338b_GLS_carmenes-eps-converted-to.jpg}. The highest peak is located at $\sim$24.4\,d followed by two other significant peaks at $\sim$8.3 and $\sim$16.6\,d. The three signals lie above FAP level 0.1\,\%~and are marked in all panels of Figs.~\ref{fig:gj338b_activity} and~\ref{fig:gj338b_GLS_carmenes-eps-converted-to.jpg}. The GLS periodogram of the CARMENES NIR RVs of GJ\,338\,B is shown in the second panel of Fig.~\ref{fig:gj338b_GLS_carmenes-eps-converted-to.jpg}. Only the peak at 16.6\,d exceeds FAP $\sim$ 10\,\%~(significance is not as high as in the optical probably because there are less NIR RVs and they have larger associated uncertainties). This signal at 16.6\,d is common to all RV and many activity index GLS periodograms of GJ\,338\,B, and as discussed in the previous Section, it is very likely related to the stellar rotation. Even though it is believed that the activity-induced signal is the strongest at blue wavelengths and is attenuated at long wavelengths in the RV time series \citep{2006ApJ...644L..75M, 2008A&A...489L...9H}, GJ\,338\,B clearly shows this peak in the near-infrared despite the larger error bars of the CARMENES NIR channel. This implies that the mechanism responsible for the activity in GJ\,338\,B is capable of imprinting its signal at both optical and near-infrared wavelengths. Figure~\ref{Fig_NIR_phase} illustrates the CARMENES NIR RVs folded in phase with the rotation period of $\sim$ 16.6\,d; the curve shows a amplitude of 4.3$\pm$1.7\,m\,s$^{-1}$.

For completeness, we also provide the GLS periodogram of the HIRES data of GJ\,338\,B in the third panel of Fig.~\ref{fig:gj338b_GLS_carmenes-eps-converted-to.jpg}. No significant signal is seen at any of the three peaks provided by the CARMENES VIS RV dataset. This is quite likely due to the insufficient cadence and long time coverage of the HIRES data. Interestingly, the combination of the HIRES and CARMENES VIS RVs recovers the 16.6\,d signal well above FAP level 0.1\,\%~ (fourth panel of Fig.~\ref{fig:gj338b_GLS_carmenes-eps-converted-to.jpg}). The second most prominent signal is located at $\sim$24.4\,d, which agrees with the results of the CARMENES VIS data. 

%%%%%%%% ALIAS

\begin{figure}[]
\includegraphics[width=\columnwidth]{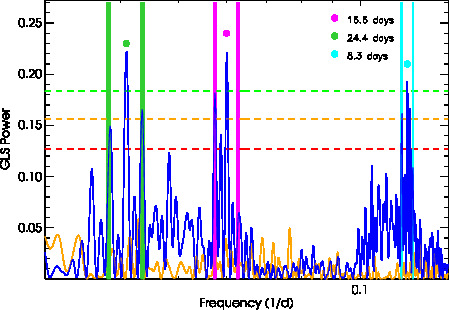}
\caption{Enlargement of the GLS periodogram of the CARMENES VIS RV data of GJ\,338\,B (blue line) around the three strongest signals (marked with color-coded dots). The window function is plotted as the orange line. Horizontal dashed lines indicate the different FAPs: 0.1\,\% (green), 1\,\% (orange) and 10\,\% (red). The 1-sidereal-year aliases around each of the strongest signals are indicated with vertical solid lines. }
\label{fig:gj338b_window}
\end{figure}

\begin{figure}[]
\includegraphics[scale=0.39]{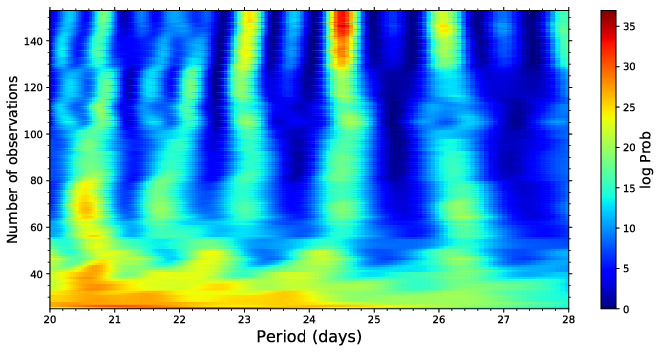}
\caption{Stacked BGLS periodogram for the CARMENES RVs data zoomed around the likely planetary orbital period of 24.4 days. The cumulative number of CARMENES observations is plotted as a function of periodicity; the color scale indicates the logarithm of the probability, where the red color stands for the most likely period. }
\label{fig:gj338b_SBGLS}
\end{figure}

In what follows, we demonstrate that the three signals ($\sim$8.3, $\sim$16.6, and $\sim$24.4\,d) of the CARMENES VIS RVs of GJ\,338\,B are not related to each other by an aliasing effect, which is typically caused by the gaps in the time coverage of the observations \citep[e.g.,][]{2010ApJ...722..937D}. To identify the presence of possible aliasing phenomena, the spectral window has to be considered. If peaks are seen in the window function, their corresponding aliases will be present in the RV periodograms as $f_{\rm alias} = f_{\rm true}\pm  mf_{\rm window}$, where $m$ is an integer, $f_{\rm true}$ is the frequency identified in the RV periodogram and $f_{\rm window}$ the frequency from the window function \citep{1975Ap&SS..36..137D}. Typical aliases are those associated with the sidereal year, synodic month, sidereal day, and solar day. The window function of the CARMENES VIS data is depicted in the top panel of Fig.~\ref{fig:gj338b_GLS_carmenes-eps-converted-to.jpg} together with the GLS periodogram of the RVs. There are two significant peaks in the window function at the sidereal year and day. Using this information, we plotted the predicted aliases corresponding to each of the three significant peaks of the CARMENES VIS RVs in Fig.~\ref{fig:gj338b_window}. From this Figure, none of the three signals are related to each other by an alias of the observing window. 

To summarize, based on the CARMENES VIS and NIR RV data we identified one strong signal at 16.61 $\pm$ 0.04\,d, which is also significant in some activity indicators, and thus we relate it to the rotation period of GJ\,338\,B. The signals at $\sim$8.3 and $\sim$24.4\,d do not have a counterpart in the activity indicators and are not aliases of the rotation signal. The $\sim$8.3-d peak is exactly at half the rotation period, and we attribute it to the first harmonic of the stellar rotation. The $\sim$24.4-d signal is not an exact multiple of the star's rotation, and we attribute it to a planetary origin. We investigate its nature in Sect. \ref{sec:Best-fit parameters}. 

To explore the stability of the $\sim$24.4-d period, we produced the stacked periodogram of the CARMENES data shown in Fig.~\ref{fig:gj338b_SBGLS}. For this purpose, we computed the Bayesian Generalized Lomb-Scargle periodogram \citep[BGLS,][]{2015A&A...573A.101M}. This formalism determines the probability between peaks of the periodogram.  The signal at 24.4\,d is very well detected with a probability above 10$^{30}$ (with the minimum probability set to 1). This signal becomes more significant by adding observations as expected for a real signal.

\begin{figure}[!h]
\centering
\includegraphics[width=\columnwidth]{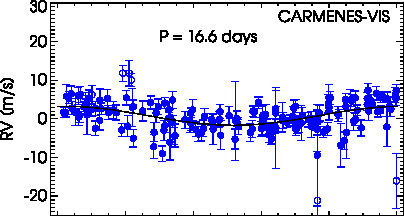}
\includegraphics[width=\columnwidth]{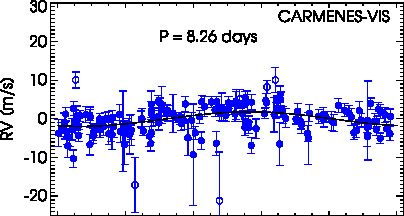}
\includegraphics[width=1.01\columnwidth]{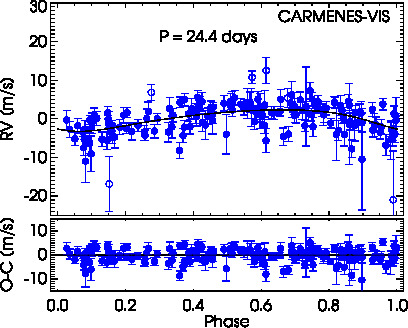}
\caption{CARMENES VIS RV data of GJ\,338\,B folded in phase with different periods. Each panel has been corrected for the others signals. {\sl Top panel:} RV data folded in phase with the stellar rotation period of 16.6\,d. Overplotted it is the best sinusoid function reproducing the data with an amplitude of 2.7$\pm$0.5\,m\,s$^{-1}$. {\sl Second panel:} RV folded in phase with the 8.3\,d signal. Another sinusoid function is fit to these data with a RV amplitude of 1.8$\pm$0.6\,m\,s$^{-1}$. {\sl Third panel:} RV data folded in phase with the planetary orbital period of 24.4\,d. The best Keplerian solution for the planet, with an RV amplitude of 2.8$\pm$0.4\,m\,s$^{-1}$, is shown with the black line. {\sl Bottom panel:} final RV residuals after removing the three signals (16.6, 8.3, and 24.4\,d) have an $rms$\,=\,3.0\,m\,s$^{-1}$. Open circles represent the RVs that deviate more than 2$\sigma$ from the trend of the original data and that were not included in the analysis but are shown in this figure.}
\label{fig:gj338b_phase_folded_prot}
\end{figure}

\subsection{Frequency content of the radial velocities of GJ\,338\,A }

The GLS periodogram of GJ\,338\,A CARMENES VIS RVs is illustrated in the top panel of Fig.~\ref{fig:gj338a_multiplot}. The highest peak is located at 16.3$^{+3.5}_{-1.3}$\,d and shows a broad structure. No other signal lies above FAP 0.1\%. As discussed in Section \ref{sec:Activity}, this periodicity is also present in the GLS peridograms of the CARMENES activity indicators (Ca\,{\sc ii} IRT, H$\alpha$, and dLW) of GJ\,338\,A and it is very likely related to the stellar rotation.

\begin{figure}[]
\centering
\includegraphics[width=\columnwidth]{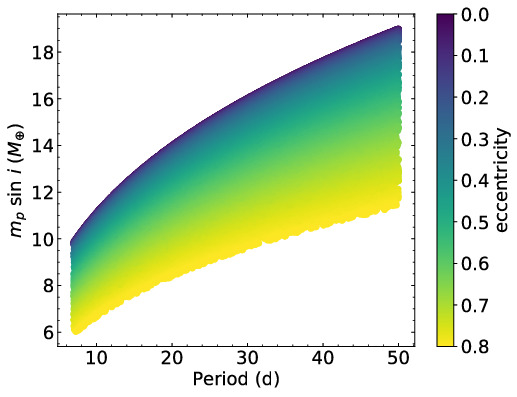}
\caption{Minimum detectable mass of any putative planet surrounding GJ\,338\,A as a function of the orbital period. We adopted the $rms$ of the CARMENES VIS RV residuals as the amplitude of the Keplerian signals. The color code stands for the different values of the eccentricity.
}

\label{fig:planet_mass_GJ338A}
\end{figure}

This activity signal can be modeled by a periodic curve with an amplitude of 5.6$\pm$0.7\,m\,s$^{-1}$ (VIS RV data). After its removal from the CARMENES VIS data, we found an $rms$ of the residuals of 4.6\,m\,s$^{-1}$ and no additional significant peak in the corresponding periodogram above a FAP of 1\,\%. We calculated the minimum mass of any putative planet around GJ\,338\,A in an orbital period range between 7 and 50\,d using the third Kepler's law, the mass of the parent star given in Table~\ref{stellar_parameters_1}, and different values of eccentricity. The minimum period is given by four times the typical Nyquist frequency of the RV time series, and the maximum period is arbitrarily chosen. Figure~\ref{fig:planet_mass_GJ338A} depicts the planetary minimum masses where we adopted, at the 1\,$\sigma$ level, the $rms$ of the CARMENES VIS RVs multiplied by a factor of 0.92 as the minimum detectable RV amplitude. \cite{2009PASP..121..365B} argued that a signal must be about 92\,\%~of the magnitude of the average residuals to be measurable through Lomb-Scargle-based periodograms. We are able to discard the presence of planets more massive than 10--19\,M$_{\oplus}$ at 7--50\,d orbits with eccentricity $\sim$0.1, respectively. For more eccentric orbits, the planetary minimum masses decreases down to 8--14\,M$_{\oplus}$ at 7--50\,d orbits.

Following the same approach as for GJ\,338\,B (see Section~\ref{sec:Best-fit parameters}), we also analyzed the GLS peridograms of the HIRES data and the combined HIRES$+$CARMENES VIS RVs of GJ\,338\,A, finding no characteristic frequency in the former and recovering the 16.3-d peak in the latter. The broad structure of the 16.3-d peak could be explained by the existence of differential rotation and/or blending of a real (very likely activity) signal at a similar period with an aliasing effect (due to one and two year aliases, time gaps between data, etc.). When removing the signal at 16.3\,d from the HIRES$+$CARMENES dataset, another signal at 17\,d became apparent above FAP 0.1\%  in the GLS periodogram of the residuals. Both 16.3\,d and 17\,d periodicities fall within the characteristic frequency interval of 15.8--26.9\,d obtained from the analysis of the photometric $V$-band data. No more significant signals are present when removing the 16.3\,d and 17\,d periodic variations. Using current data, we cannot discard the presence of super-Earth or less massive planets ``hidden" in the broad 16.3\,d peak of the RV periodogram.

%--------------------------------------------------------------------
%--------------------------------------------------------------------
%--------------------------------------------------------------------
%--------------------------------------------------------------------
%--------------------------------------------------------------------
%--------------------------------------------------------------------
%--------------------------------------------------------------------

\begin{table}[t]
\begin{small}
\centering
\caption{Comparison of different solutions for GJ\,338\,B.}
\label{tab:gj338b_model_comparison}
\begin{tabular}{l l c c c}

\hline
\hline
\noalign{\smallskip}
Model & Prior & GP kernel & $\ln \mathcal{L}$ & BIC$^{(1)}$\\
\noalign{\smallskip}	
\hline	
\noalign{\smallskip}
BM$^{(2)}$ &  $\mathcal{U}_{\rm offset}$(-100, 100) & ... & -569 & 1153\\
								 & $\mathcal{U}_{RV_{\rm jitter}}$(-10, 2) &  & &\\
								&  $\mathcal{U}_{\rm trend}$(-1, 1) &  & &\\
BM+GP				    & $\mathcal{U}_{\omega_0}$(-10, 5)   & SHO & -547 & 1124\\
							    &  $\mathcal{U}_{Q}$(-10, 5) &  &  &\\
							    & $\mathcal{U}_{S_0}$(-10, 5) &  & &\\
BM+1pl      			    & $\mathcal{U}_{\rm pl}$ (24.39, 24.5) &     ...      & -546 & 1132\\
BM+GP+1pl			& $\mathcal{U}_{\rm pl}$ (24.39, 24.5)  & SHO & -517 & 1089\\
BM+2GP+1pl   		&$\mathcal{U}_{\rm pl}$ (24.39, 24.5)   & SHO & -513 & 1096\\
BM+GP+2pl		    & $\mathcal{U}_{\rm pl_{1}}$ (24.39, 24.5)  & SHO & -497 & 1075\\
								& $\mathcal{U}_{\rm pl_{2}}$ (8.3, 8.4)  & 			& & \\

\noalign{\smallskip}	
\hline	
\noalign{\smallskip}
\end{tabular}
\tablefoot{$^{(1)}$ BIC corresponds to the Bayesian Information Criterion. $^{(2)}$ BM stands for the base model containing offsets, RV jitter, and a linear trend. }
\end{small}
\end{table}

%%%%%%%%%%%%%%%%%%%%%%%%%%%%%%%%%%%%%%%%%%%%%

%%%%%%%%%%%%%%%%%%%%%%%%%%%%
\begin{table*}[t]
\centering
\caption{Keplerian orbital parameters of GJ\,338\,Bb from pre-whitening and Gaussian process regresion methods .}
\label{tab:gj338b_from_GP_and_RV}
\begin{tabular}{l c c c}

\hline
\hline
\noalign{\smallskip}
                   & Pre-whitening  & GP regression   &    \\
\noalign{\smallskip}	
\hline	
\noalign{\smallskip}
Parameter & 	GJ\,338B b 		&	Prior 					& GJ\,338B b \\
\noalign{\smallskip}	
\hline	
\noalign{\smallskip}
$P$  (d) 								& $24.40 \pm 0.04$ & $\mathcal{U}$ (24.39, 24.50) 		&$24.45 \pm 0.02$ \\
$T_0$ (BJD-2,400,000)$^{(1)}$ & $57831.2 \pm 2.0$ & $\mathcal{U}$ (0, 30 )		&$57517.06^{+6.15}_{-4.72} $ \\
$e$ 									& $0.25 \pm 0.13$ & $\mathcal{U}$ (0, 0.8 )		& $0.11^{+0.11}_{-0.08} $\\
$\omega$ (deg) 				& $144.9 \pm 31.8$  &	$\mathcal{U}$ (-2$\pi$, 2$\pi$)		& $204.3^{+94.4}_{-70.0}$ \\
$K$ (m/s)							& $2.8 \pm 0.4$     &  $\mathcal{U}$ (0, 10 )  	     & $3.07 \pm 0.37$ \\
$v_0$ (m/s)$^{(2)}$ 		& $0.09 \pm 0.23$ & 	$\mathcal{U}$ (-100, 100 ) & $3.16 \pm 0.32$\\
\noalign{\smallskip}	
\hline	
\noalign{\smallskip}
$m_{\rm p}\sin i$ ($\rm M_{ \oplus}$)$^{(3)}$ & $9.15 \pm 1.11$ & & $10.27^{+1.47}_{-1.38}$ \\
$m_{\rm p}\sin i$ ($\rm M_{ \oplus}$)$^{(4)}$ & $8.90 \pm 1.20$ & &   $9.97^{+1.47}_{-1.38}$ \\
%$m_{\rm p}\sin i^{(2)}$  ($\rm M_{Jup}$) 	  & &$0.029 \pm 0.004$ \\
$a$ (au) 															 & $0.142 \pm 0.014$ & &$0.141 \pm 0.005$ \\
$T_{\rm eq}$ (K)$^{(5)}$								&	 300--390~$\pm$~30   & &  301--391~$\pm$~20 \\
\noalign{\smallskip}
\hline
\end{tabular}
\tablefoot{$^{(1)}$ $T_0$ corresponds to the periastron passage. The reference time for the Gaussian process (GP) regression is 2457500.0. $^{(2)}$ Arbitrary zero point applied to CARMENES VIS RVs. $^{(3)}$ Derived by adopting the mass of $0.64 \pm 0.07$\,$\rm M_{\odot}$ for the parent star. $^{(4)}$ Derived with stellar mass determined by \cite{2019A&A...625A..68S}. $^{(5)}$\,For Bond albedo in the interval 0.65--0.0. }
\end{table*}
%%%%%%%%%%%%%%%%%%%%%%%%%%%%

\section{Planet orbiting GJ\,338\,B}
\label{sec:Best-fit parameters}

\subsection{Pre-whitening method}

We obtained the best-fit orbital parameters of the planet at $\sim$24.4\,d using the {\tt RVLIN} code\footnote{The {\tt RVLIN} code is available at {\tt http://exoplanets.org/code/}} \citep[][]{2009ApJS..182..205W}. This code is based on the Levenberg-Marquardt \citep{Levenberg44,Marquardt63} algorithm and contains a set of routines that fits an arbitrary number of Keplerian curves to RV data. Our approach was as follows: first, we removed the ``stellar contribution'', that is, activity, with a rotational period of 16.6\,d from the observations by fitting a sinusoid function to the CARMENES VIS data. The fit, shown in the upper panel of Fig.~\ref{fig:gj338b_phase_folded_prot}, yielded an RV amplitude of 2.7\,m\,s$^{-1}$ (smaller than that obtained from the analysis of the NIR RVs, see Section~\ref{Radial velocity analysis}). The residuals had an $rms$ of 3.9\,m\,s$^{-1}$, which was slightly smaller than the initial standard deviation of the CARMENES VIS data, but still larger than the typical uncertainty associated with the individual RV measurements.

A new GLS periodogram of the residuals was obtained (it is depicted in panel (e) of Fig.~\ref{fig:gj338b_GLS_carmenes-eps-converted-to.jpg}). The planetary signal at $\sim$24.4\,d, which was also obvious in the original GLS periodogram, is now enhanced. The $\sim$8.3\,d peak still remains, although its significance is at the FAP level 10\,\%. Because it is likely related to stellar activity, we removed its contribution from the VIS RV data by fitting a second sinusoid with a period of 8.26\,d and and RV amplitude of 1.8\,m/s (middle upper panel of Fig.~\ref{fig:gj338b_phase_folded_prot}). The new residuals have an $rms$ of 3.6\,m\,s$^{-1}$. With these subsequent sinudoidal subtractions from the CARMENES VIS RV data, the only significal signal of the residuals has a characteristic period of 24.4\,d (panel (f) of  Fig.~\ref{fig:gj338b_GLS_carmenes-eps-converted-to.jpg}).

With the RVLIN code, we found the best Keplerian solution, which is illustrated in the third panel of Fig.~\ref{fig:gj338b_phase_folded_prot}, where the RV data are folded in phase with the planetary orbital period. The CARMENES VIS RV measurements that were not included in the analysis (open symbols in the RV plots) follow the trend and lie at the expected phases. The corresponding best-fit orbital parameters and their error bars are reported in Table~\ref{tab:gj338b_from_GP_and_RV}. The best-fit orbital period is $P$\,=\,24.40\,$\pm$\,0.04\,d, and the planet has a minimum mass of 9.2\,$\pm$\,1.1\,M$_{\rm \oplus}$ (computed by adopting the mass of 0.64\,$\pm$\,0.07\,M$_\odot$ for the parent star, as determined here). This planetary minimum mass would be 8.9$\pm$1.2\,M$_\oplus$ if the stellar mass determined by \citet{2019A&A...625A..68S} were adopted instead. The planet candidate GJ\,338\,Bb is located at an orbital separation of 0.142\,$\pm$\,0.014\,au from its parent M0-type star, and despite its proximity, the planetary orbit appears to be slightly eccentric ($e = 0.25 \pm 0.13$), although given the error bar of the eccentricity, the orbit is also compatible with an almost circular orbit. The amplitude of the planet RV curve is 2.8\,$\pm$\,0.4\,m\,s$^{-1}$, about 38\,\%~smaller than the dispersion of the residuals data and similar to the mean error bar of individual CARMENES VIS RV data. The RV amplitudes of the stellar activity and the planet-induced signal are quite similar.

The RV residuals obtained after removing the stellar and planetary contribution from the CARMENES VIS RV data are shown in the bottom panel of Fig.~\ref{fig:gj338b_phase_folded_prot}. They have an $rms$ of 3.0\,m\,s$^{-1}$, still larger than the mean error bar of individual RV measurements, but similar to the typical $rms$ average of CARMENES data. The GLS periodogram of these residuals is displayed in panel (g) of Fig.~\ref{fig:gj338b_GLS_carmenes-eps-converted-to.jpg}; it shows no other significant peak. The results are not modified by changing the order of the pre-whitening method, thus giving robustness to the discovery of the small-mass planet around GJ\,338\,B. Table\ref{tab:gj338b_from_GP_and_RV} summarises the orbtial parameters.

For completeness, we combined the CARMENES VIS and HIRES RV datasets of GJ\,338\,B. The fourth panel of Fig.~\ref{fig:gj338b_GLS_carmenes-eps-converted-to.jpg} illustrates the corresponding GLS periodogram. We performed the same pre-whitening analysis as previously described. The amplitude of the stellar activity of the combined RV curve was determined at 2.7\,m\,s$^{-1}$, which fully agrees with that from the CARMENES-data only. The first residuals have an $rms$ of 4.1\,m\,s$^{-1}$, which is indeed slightly higher than the first residuals of the CARMENES VIS RV data only. However, from this point of the analysis to the end of the process, the noise added by the HIRES data and the poor temporal coverage of this dataset prevented us from finding a consistent orbital solution ($P=24.40$\,d) despite the fact that the periodograms of the combined HIRES$+$CARMENES VIS RV residuals (not shown here) do not differ significantly from Fig.~\ref{fig:gj338b_GLS_carmenes-eps-converted-to.jpg}. The eccentricity value that we found for the planet orbital solution with the combined CARMENES VIS and HIRES RV data sets is $e=0.1$. The lessons learnt from this exercise is that the HIRES RVs do not contradict the findings of the CARMENES data and that the eccentricity of the orbit of the planet around GJ\,338\,B is not well constrained with the current data.

%--------------------------------------------------------------------
%--------------------------------------------------------------------
%--------------------------------------------------------------------
%--------------------------------------------------------------------
%--------------------------------------------------------------------
%--------------------------------------------------------------------
%--------------------------------------------------------------------

\begin{figure*}[!ht]
\centering
\begin{minipage}{0.49\linewidth}
\includegraphics[angle=0,scale=0.35]{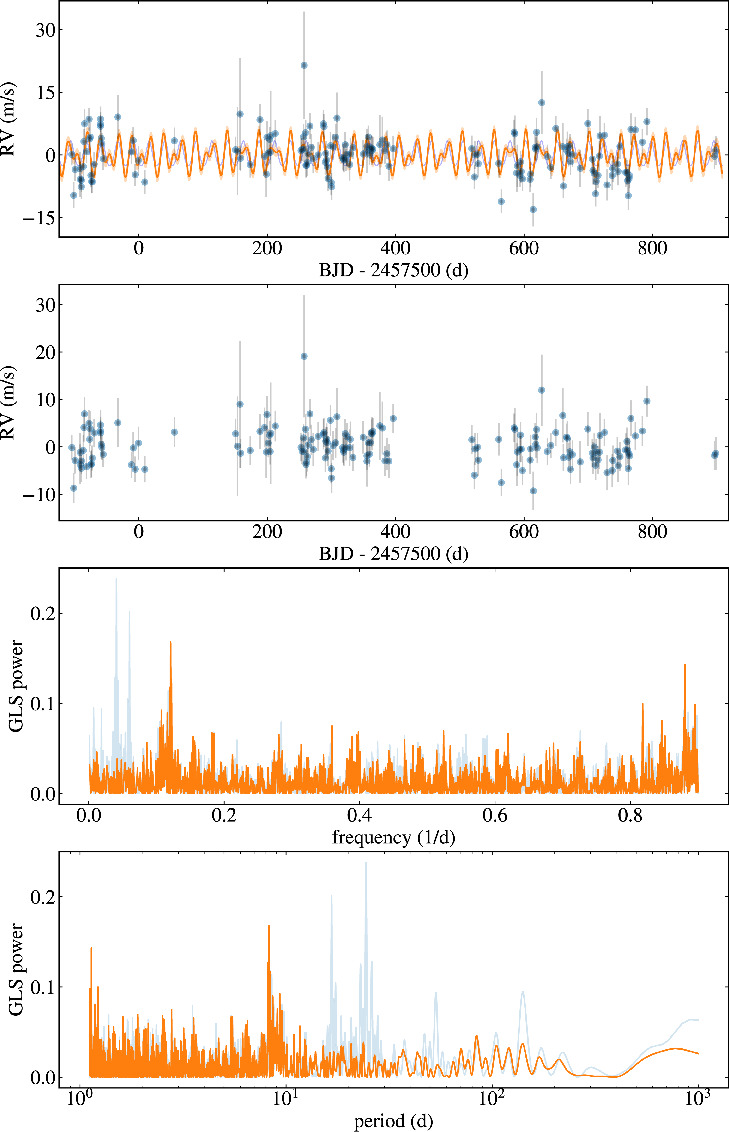}
\end{minipage}
\begin{minipage}{0.49\linewidth}
\includegraphics[angle=0,scale=0.35]{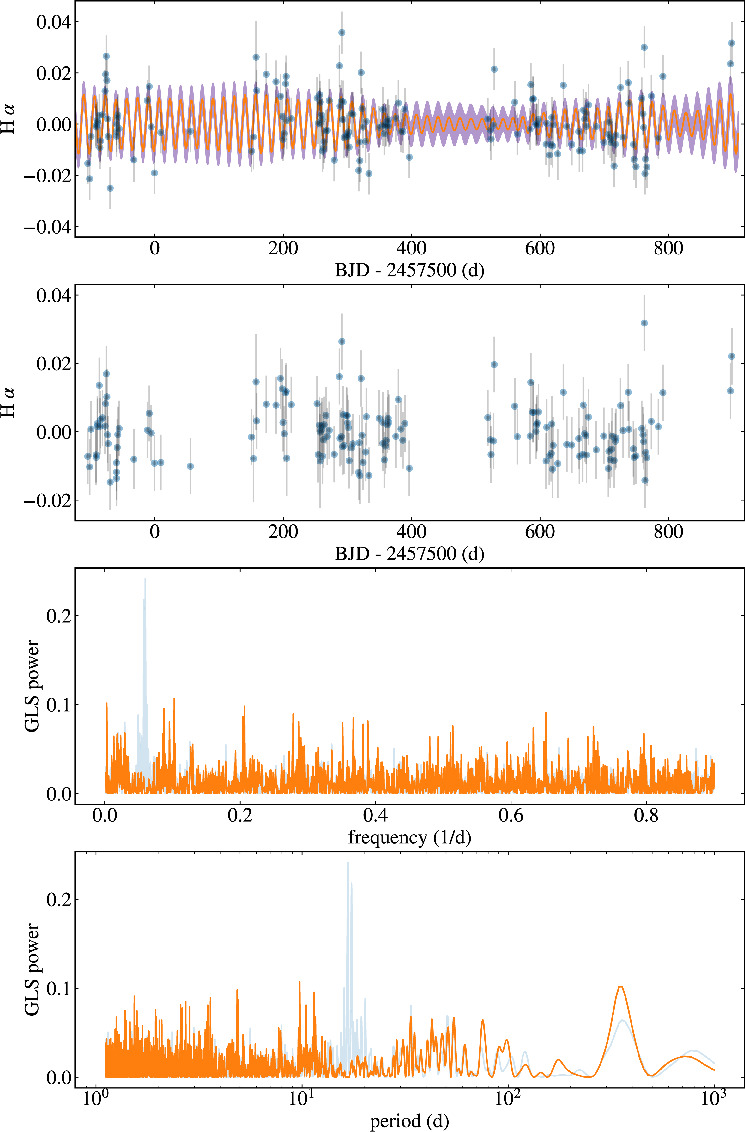}
\end{minipage}
\caption{{\sl Left panels:} top panel illustrates the CARMENES VIS RVs (blue dots) together with a Keplerian model combined with a simple-harmonic oscillator (SHO) model shown in orange. The Keplerian-only curve is shown with a light blue color. The second panel depicts the RV residuals after subtracting the combined Keplerian $+$ simple-harmonic oscillator model from the original data. The third and bottom panels show the GLS periodograms of the original data (light blue line) and of the RV residuals (orange) as a function of frequency and period. The model accounts for the two strongest peaks at 16.6\,d and 24.4\,d. 
{\sl Right panels:} CARMENES H$\alpha$ data compared to a simple harmonic oscillator (SHO) model (see Section \ref{subsec:GP model}) and the residuals (top two panels), as well as the periodogram as a function of frequency and period (lower two panels). In the top panel, the best fit model is shown in orange, the 1$\sigma$ uncertainty in blue. The periodograms show the data in light blue and the residuals in orange. 
}
\label{Fig_GPRV}
\end{figure*}

\begin{figure}[]
\centering
\includegraphics[width=\columnwidth]{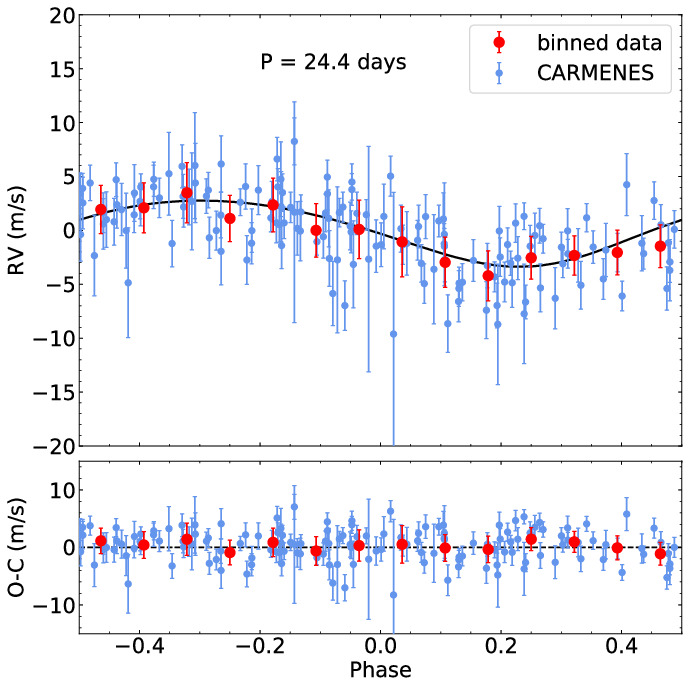}
\caption{{\sl Top panel:} planetary signal of GJ\,338\,B CARMENES VIS RVs (blue dots) folded in phase with the orbital period derived from the GP regression analysis (the base model and the stellar activity were removed). The best Keplerian solution (black line) has an RV amplitude of 3.07$\pm$0.37\,m\,s$^{-1}$. The red dots correspond to the binned data. {\sl Bottom panel:} RV residuals after the subtraction of the model. The weighted $rms$ of the residuals is 3.2\,m\,s$^{-1}$. 
}
\label{fig:phase_GP_data}
\end{figure}

\subsection{Gaussian process regression method}
\label{subsec:GP model}

We also followed a different approach to study in detail the CARMENES VIS RV curve of GJ\,338\,B. This approach has the advantage of simultaneously fitting the stellar activity and the planetary signal(s). The match between the dominant period in dLW and H$\alpha$ indices (Section~\ref{sec:Activity}) with one of the strongest signals in the RVs (Section~\ref{Radial velocity analysis}) motivated a simultaneous fit of these activity indicators together with the RVs in order to account for the correlated RV noise. We used the {\tt celerite} package \citep{2017AJ....154..220F} and its damped simple harmonic oscillator (SHO) kernel of the form
\begin{equation}
S(\omega) = \sqrt{\frac{2}{\pi}} \frac{S_0\,\,\omega_0^{4}}{(\omega^2 - \omega_0^2)^2 + \omega_0^2 \omega^2 / Q^2}
\end{equation}
The parameters of the model are the oscillator eigen-frequency ($\omega_0$), the quality factor ($Q$, corresponding to the inverse of the damping time scale), and the amplitude of the damped oscillator ($S_0$). The former two are constrained by the activity indicator and an SHO Gaussian process (GP) kernel with these parameters, but with its own amplitude, is added to the planetary model. In addition to this, we used a model with individual offsets and RV jitter, as well as a linear trend, as the base line model (BM). The RV data and the activity indicators were then solved at the same time.

We used the Bayesian Information Criterion \citep[BIC,][]{2007MNRAS.377L..74L} to identify the optimal solution. When fitting models, it is possible to increase the log-likelihood (ln\,$\mathcal{L}$) by adding parameters, but the model may result in overfitting. The BIC resolves this problem by introducing a penalty term of the number of parameters in the model. It can be mathematically defined as follows:
\begin{equation}
{\rm BIC} = k\, \ln(n) - 2\,\ln(\mathcal{L}),
\end{equation}
where $n$ is the number of data points, $k$ the number of free parameters to be estimated, and $\mathcal{L}$ the maximized value of the likelihood function of the model. The lower the BIC value, the better the model is.

In a first approach, we modeled the base model together with one SHO GP kernel (which picks the 16.6-d activity simultaneously in the RVs and in H$\alpha$ or dLW), and the base model plus one planet with an orbital period of 24.4 d. With respect to the base model, these two models represented an improvement of the BIC value. Next, we explored how to lower the BIC parameter even further.

In the second approach, we modeled the base model, one Keplerian signal at 24.4\,d, and one SHO GP kernel to account for the stellar activity with a cycle of 16.6\,d. The fitted stochastic oscillator has a very high quality factor, that is, the life time is of the order of the duration of the observation. With this high quality factor, the GP kernel only acts in a very limited period/frequency range around its base frequency. The signal at 8.3\,d (=1/2 $P_{\rm rot}$) is not removed. The planetary parameters of amplitude and orbital periodicity remained unaffected by the GP model. 

In a third approach, we assumed that the peaks at 8.3\,d and 16.3\,d are due to stellar rotation and the peak at 24.4\,d is originated by a planet. We run the base model and added the planet and two SHO kernels that correspond  to the stellar rotation period and its first harmonic ($P_{\rm rot}$/2). We forced the damping time of the two SHO kernels to be identical. The derived planetary parameters are compatible with the findings of previous models within the quoted error bars.

Our last approach consisted in treating the 8.3\,d ($P_{\rm rot}$/2) signal as an additional Keplerian signal. This is, the base model, two planets at 8.3\,and 24.4\,d, and the stellar activity modeled with a SHO GP kernel at 16.6\,d. Also this time, the planetary parameters of the 24.4-d planet remain unaffected. All models, including their priors, log-likelihood and BIC values, are summarized in Table~\ref{tab:gj338b_model_comparison}.

According to the BIC criterion, the best solution for the CARMENES VIS RV curve of GJ\,338\,B is given by the base model, a SHO GP kernel modeling the stellar activity at 16.6\,d, and two planets with orbital periods of 8.3\,d and 24.4\,d (BM+GP+2pl in Table~\ref{tab:gj338b_model_comparison}). The orbital parameters of the outermost planet are listed in Table~\ref{tab:gj338b_from_GP_and_RV}. The 8.3-d planet would have the following derived parameters: minimum mass of 4.3\,$\pm$\,1.5 M$_\oplus$ and semi-major axis of 0.067\,$\pm$\,0.015\,au. However, as mentioned in Section~\ref{Radial velocity analysis}, the 8.3-d period is exactly half of the stellar rotation cycle; with the current data we cannot confirm the presence of this small planet around  GJ\,338\,B. Therefore, we adopted the model given by thethe second approach, this is the base model, one SHO kernel for the stellar activity at the star's rotation, and the planet at 24.4\,d (BM+GP+1pl in Table~\ref{tab:gj338b_model_comparison}). This is the solution for the GJ\,338\,B planetary system that we follow in this paper.

Summarizing, the eigen-frequency and the quality factor are converted to period and damping time, the uncertainties are derived from an MCMC procedure using {\tt emcee} \citep{2013PASP..125..306F} with 200 walkers and 10,000 iterations after the burn-in phase. The autocorrelation length averaged over the walkers is 550, and the acceptance rate is 22\,\%~and during burn-in it is 35\,\% (as described by \citealt{2013PASP..125..306F}). We also checked the chains of each parameter as a function of the step and after the burn-in phase, and the walkers were wandering around the best parameter. The MCMC posterior distributions for the fitted parameters are presented in Fig. \ref{fig:corner_plot_GP-eps-converted-to.jpg}. The derived orbital parameters using the GP regression are reported in Table~\ref{tab:gj338b_from_GP_and_RV} and they are compatible within the quoted error bars with the values obtained in the previous section using the pre-whitening approach and the RVLIN code. 

In the top-left panel of Fig.~\ref{Fig_GPRV}, the solution combining stellar activity (as measured from the H$\alpha$ index, see right panels of Fig.~\ref{Fig_GPRV}) and the planetary signal is depicted as a function of time (with an arbitrary zero date). The stellar activity was modeled with a damped oscillator with a period of 16.6\,d and a long damping time scale of about 250\,d. The weighted $rms$ of the CARMENES VIS RVs after model subtraction is 3.2\,m\,s$^{-1}$, and 0.0084 for the H$\alpha$ index. The CARMENES VIS RV residuals are illustrated in the second panel of Fig.~\ref{Fig_GPRV}, while the third and four panels stand for the GLS periodograms of the RV original and residual data as a function of frequency and period. The RV data and the residuals folded in phase at the best Keplerian solution obtained with the GP regression are illustrated in Fig.~\ref{fig:phase_GP_data}. The RV amplitude of Keplerian signal is 3.07$\pm$0.37\,m\,s$^{-1}$.

%-----------------------------

%--------------------------------------------------------------------
%--------------------------------------------------------------------
%--------------------------------------------------------------------
%--------------------------------------------------------------------
%--------------------------------------------------------------------
%--------------------------------------------------------------------

\section{Discussion}
\label{Discussion}

The two stars are very similar in mass, spectral type, brightness, and even rotation period, but they appear to differ in the architecture of their planetary systems. Given the rotation period of GJ\,338\,B ($16.61 \pm 0.04$\,d), its projected rotational velocity ($v \sin i = 2.3 \pm 1.5$\,km\,s$^{-1}$), and the star’s radius ($0.58 \pm 0.03$\,$R_{\odot}$) as taken from the literature, it is not possible to determine a precise inclination angle of the stellar rotation angle. It is likely that the projected rotational velocity is smaller than the values published in the literature. The rotation period agrees with the observed relation between X-ray luminosity and stellar activity or rotation from, for example, \cite{2011ApJ...743...48W}.

Here, we have used two mass determinations for the M0V-type star GJ\,338\,B. One was obtained from the most likely astrometric solution of the stellar binary combined with the mass ratio determination from absolute radial velocity measurements. This value (Table~\ref{tab:gj338_orbital_param}) is fully independent of any evolutionary model and mass-luminosity calibrations. Despite the degeneracy of the astrometric solution (Section~\ref{Orbital_solution}), the individual masses of the stellar components are well constrained. The second measurement provided by \cite{2019A&A...625A..68S} is based on various well known calibrations that are valid for low-mass stars. The two determinations differ by less than 10\,\%, thus having little impact in the mass determination of the planet orbiting this star.

\begin{figure}[]
\centering
\includegraphics[width=\columnwidth]{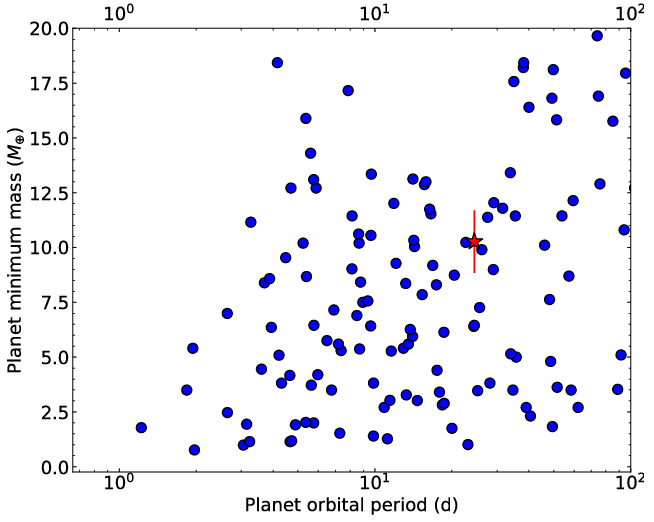}
\includegraphics[width=\columnwidth]{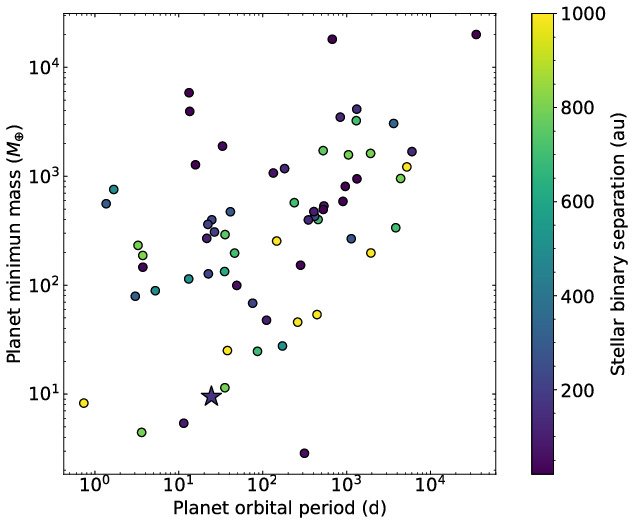}
\caption {{\sl Top panel:} planetary minimum mass versus orbital period for known Neptune-sized and super-Earth planets (blue dots) around M dwarfs. The planet GJ\,338\,Bb is plotted as the red star symbol.
% (http://www. exoplanet.eu – black dots, 2018 August 9)
{\sl Bottom panel:} planetary minimum mass versus semi-major axis for all known planets (color dots) in S-type configuration around binary systems, where the stellar binaries have projected physical separation $s \le$ 1000\,au (color bar). The stellar binary separation between the A and B members of our binary system has a semi-major axis of 130.9\,au and is plotted as star symbol.
 }
\label{fig:planets_around_dM}
\end{figure}

The top panel of Fig.~\ref{fig:planets_around_dM} shows the location of GJ\,338\,Bb  in the planetary minimum mass versus planetary orbital period diagram together with other Neptune-size and super-Earth planets known around M-type dwarfs. Our planet does not stand out in this diagram, neither it does when we select only those planets that are orbiting one of the stars of the binary systems \citep[S-type configuration;][]{1988A&A...191..385R}. The bottom panel of Fig.~\ref{fig:planets_around_dM} displays such comparison, where we included the $\sim$70 discovered planets in S-type configuration with stellar projected physical separations $s \le$\,1000\,au and known planet orbital periods and minimum masses. The majority of these planets have been identified around stars in binary systems with large physical separation, typically mean values $s >$ 400\,au. GJ\,338\,Bb is one of the least massive planets identified in one component of a stellar binary with separation well below of 400\,au. The super-Earth planet around GJ 338 B adds to the increasing list of planets orbiting one of the stars of stellar binary systems. We remark the similarity of the two stellar components of the GJ\,338 binary and also provide various existing examples with very similar components such as: WASP-2\,AB, XO-2N, XO-2S, HD\,177830\,AB, Kepler\,296\,AB, WASP-94\,AB and Gliese\,15\,AB \citep[][]{2007MNRAS.375..951C, 2008ApJ...677.1324T, 2011ApJ...727..117M, 2014A&A...567L...6D, 2014ApJ...784...45R, 2014A&A...572A..49N, 2018A&A...617A.104P}.

Following commonly used expressions \citep[e.g.,][]{1984Icar...58..121B, 1999ASIC..532..189S,2013PASP..125..933S}, we derived the probability that GJ\,338\,Bb transits in front of the disk of its parent star to be 5$^{+2}_{-1}$\,\%. Using the unbiased forecasting model presented by \cite{2017ApJ...834...17C}, we predict the planetary radius to be $3.13^{+1.37}_{-0.90}\,\rm R_{\oplus}$; with this, the depth of any putative transit signal would be 0.24$^{+0.05}_{-0.01}$\,\%~as computed for a planetary mass of $10.27^{+1.47}_{-1.38}$\,$\rm M_{\oplus}$. \textit{TESS} observed the GJ\,338 binary system in sector 21 between 2020 January 21 and February 18. Due to the pixel scale of TESS (21''/pixel), the two stars GJ\,338\,A and~B are not resolved.

We adopted the traditionally defined habitable zone as the circumstellar region in which a terrestial-mass planet with a CO$_2$--H$_2$O--NO$_2$ atmosphere can sustain liquid water on its surface \citep{2007A&A...476.1373S}. We estimated the theoretical equilibrium temperature of GJ\,338\,Bb by using the Stefan--Boltzmann equation, the stellar parameters of Table~1, and two extreme values of the albedo ($A$ = 0.0 and 0.65; the highest albedo corresponds to the case of Venus in our Solar System). The results are $T_{\rm eq}$ = 391\,$\pm$\,20\,K for a non-reflecting planet and 301\,$\pm$\,20\,K for the high-reflectance planet. The error bars come from the uncertainties in the stellar luminosity and semi-major axis of the planet. \cite{2013ApJ...765..131K} calculated conservative and optimistic estimates of the habitable zone (water-loss and maximum greenhouse limits); by using their determinations of the boundaries of the habitable zone, we found them to be in the interval 0.29--0.57\,au around GJ\,338\,B. This implies that the planet GJ\,338\,Bb lies inside the inner boundary of the habitable zone which is consistent with the hot equilibrium temperature estimated above. Previous works \citep[e.g.,][]{1993Icar..101..108K,2012MNRAS.427.2239R,2017CeMDA.129..509B} established that low-mass stellar hosts can induce strong tidal effects on potentially habitable planets since they tend to be on close-in orbits. True planet habitability may be compromised for planets orbiting within the tidal lock radius of the star. Figure 16 reported by \cite{1993Icar..101..108K} shows the tidal lock radius for various stellar spectral classes, including M0\,V. From this Figure, it becomes apparent that GJ\,338\,Bb is very likely tidally locked given its short orbit.

%--------------------------------------------------------------------
%--------------------------------------------------------------------
%--------------------------------------------------------------------
%--------------------------------------------------------------------
%--------------------------------------------------------------------
%--------------------------------------------------------------------
%--------------------------------------------------------------------

\section{Conclusions}
\label{Summary and conclusions}

With the CARMENES spectrograph, we monitored the very bright M0.0\,V stars GJ\,338\,B (HD 79211) and GJ\,338\,A (HD 79210), which form a wide binary in the nearby solar vicinity. We also provided a refined astrometric and spectroscopic solution of the stellar binary orbit by using all data from the literature available to us. The adopted Keplerian solution for the two stars yielded an orbital period of 1295\,a and a semi-major axis of 130.9\,au. The orbital period value reported here is significantly larger than those previously given in the literature (e.g., \citealt{1972AJ.....77..759C}). Despite the great similarity between the two stars, we determined slightly different mass values for the A and B stellar components (0.69$\pm$0.07 and 0.64$\pm$0.07\,M$_\odot$, respectively), suggesting that the A stellar member is about 7\%~more massive than the B component. Furthermore, they are also consistent with the masses derived by \cite{2019A&A...625A..68S}.

All activity indicators provided by CARMENES VIS data (Ca\,{\sc ii} infrared triplet, H${\alpha}$ line, chromatic spectral index, the differential line width and cross correlation function analysis), together with the new photometry acquired by our group at LCO and SNO, indicate that GJ\,338\,B and A are active stars and rotate with periodicities of $P$ = 16.61$\pm$0.04\,d and $P$ = 16.3$^{+3.5}_{-1.3}$\,d, respectively, which agrees with the expected X-ray luminosity--activity relations. 

The detailed analysis of CARMENES RVs led to the discovery of a super-Earth planet orbiting GJ\,338\,B with an orbital period of 24.45\,$\pm$\,0.02\,d, and a minimum mass of 10.27$^{+1.47}_{-1.38}$\,$M_{\oplus}$, and a semi-major axis of 0.141\,$\pm$\,0.005\,au. Another period was found in the CARMENES RV data at $\sim$\,8.3\,d, but without sufficient number of arguments to prove its Keplerian nature and because it lies at half of the stellar rotation period, we attributed it to the first harmonic of the stellar activity. A related analysis concluded that GJ\,338\,A does not have planets of similar mass or more massive than GJ\,338\,Bb. The majority of the stellar binary systems with planet hosts have larger separations than the pair GJ\,338\,A and~B and their planets are also larger in mass than GJ\,338\,Bb. Therefore, GJ\,338\,Bb has become one of the least massive planets ever discovered in one star of a binary system of relatively small separation.

%---------------------------------------------------------------

\begin{acknowledgements}

We wish to thank the anonymous referee for helpful comments and suggestions, which helped to improve the manuscript. We are grateful to Prof. B. D. Mason for providing Washington Double Star astrometric data for the stellar binary. Based on observations collected at the German-Spanish Astronomical Center, Calar Alto, jointly operated by the Junta de Andalauc\'ia and the Instituto de Astrof\'isica de Andaluc\'ia (CSIC), and observations from Las Cumbres Observatory Global Telescope (LCOGT) network. LCOGT observations were partially acquired via program number TAU2019A-002 of the Wise Observatory, Tel-Aviv University, Israel. CARMENES is an instrument for the Centro Astron\'omico Hispano-Alem\'an de Calar Alto (CAHA, Almer\'ia, Spain). CARMENES is funded by the German Max-Planck- Gesellschaft (MPG), the Spanish Consejo Superior de Investigaciones Cient\'ificas (CSIC), the European Union through FEDER/ERF funds, and the members of the CARMENES Consortium (Max-Planck-Institut f\"ur Astronomie, Instituto de Astrof\'isica de Andaluc\'ia, Landessternwarte Ko\"onigstuhl, Institut de Ci\` encies de l'Espai, Insitut f\"ur Astrophysik G\"ottingen, Universidad, Complutense de Madrid, Th\"uringer Landessternwarte Tautenburg, Instituto de Astrof\'isica de Canarias, Hamburger Sternwarte, Centro de Astrobiolog\'ia and Centro Astron\'omico Hispano-Alem\'an), with additional contributions by the Spanish Ministry of Economy, the state of Baden-W\"uttemberg, the German Science Foundation (DFG), the Klaus Tschira Foundation (KTS), and by the Junta de Andaluc\'ia. This work is supported by the Spanish Ministery for Science, Innovation, and Universities through projects AYA-2016-79425-C3-1/2/3-P, AYA2015-69350-C3-2-P, ESP2017-87676-C5-2-R, ESP2017-87143-R. The Instituto de Astrof\'\i sica de Andaluc\'\i a is a Centre of Excellence `Severo Ochoa'' (SEV-2017-0709). The Centro de Astrobiolog\'ia (CAB, CSIC-INTA) is a Center of Excellence ``Maria de Maeztu''.

\end{acknowledgements}


\bibliographystyle{aa} % style aa.bst
\bibliography{bibliography.bib} % your references Yourfile.bib

%--------------------------------------------------------------------
%--------------------------------------------------------------------
%--------------------------------------------------------------------
%--------------------------------------------------------------------
%--------------------------------------------------------------------
%--------------------------------------------------------------------

%\newpage
%----------------------------------------------------------------------------------------
%	APPENDIX
%----------------------------------------------------------------------------------------

\begin{appendix} %First appendix

\section{Figures}

\begin{figure*}[]
\includegraphics[scale=0.33]{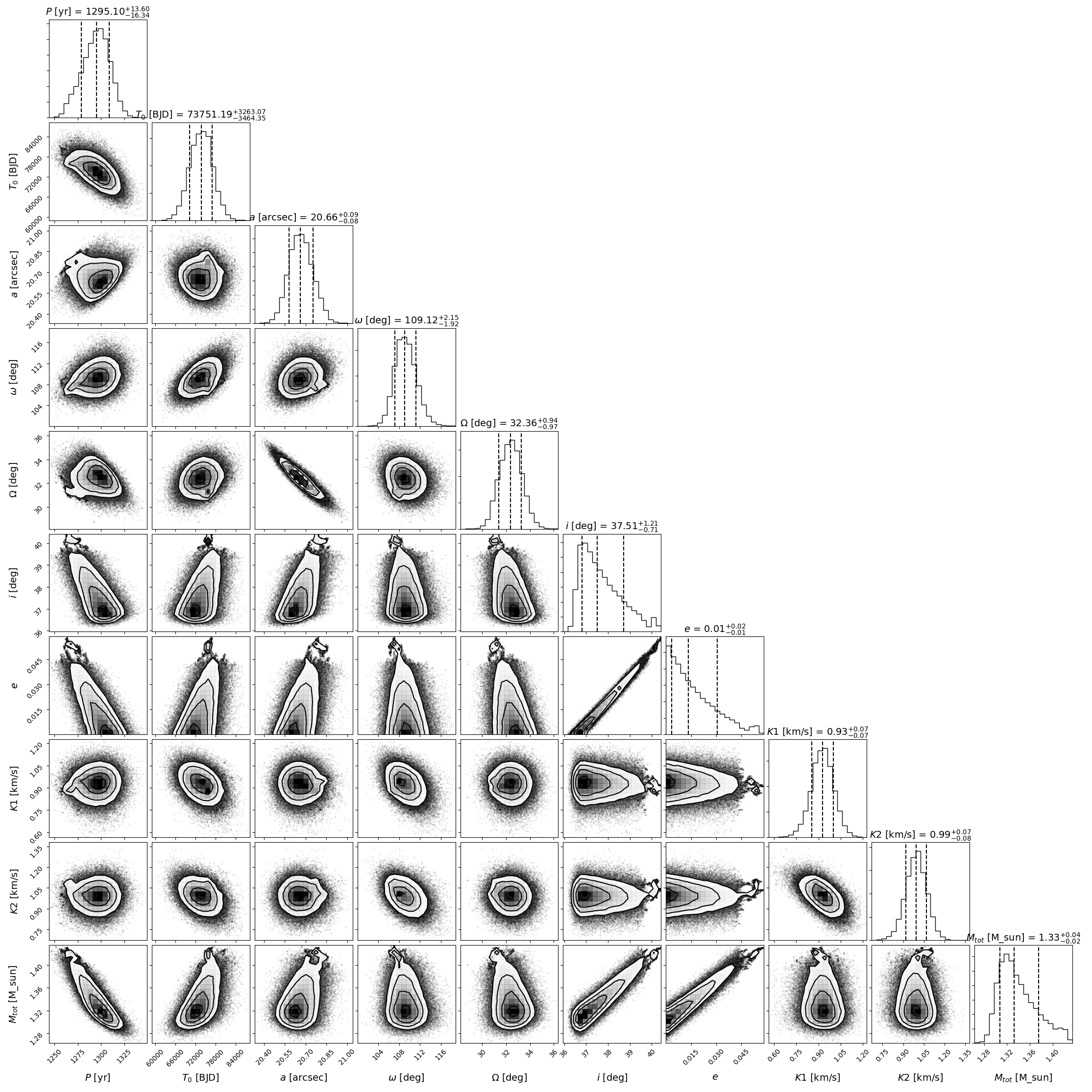}
\caption{MCMC distribution of the combined astrometric and spectroscopic analysis of the GJ 338 AB binary system. The vertical dashed lines indicate the 16\%, 50\%, and 84\% quantiles of the fitted parameters; this corresponds to 1\,$\sigma$ uncertainty. The reference time of $T_0$ distribution is 2,200,000.
}
\label{fig:my_corner_output.png}
\end{figure*}

\begin{figure*}[]
\includegraphics[scale=0.47]{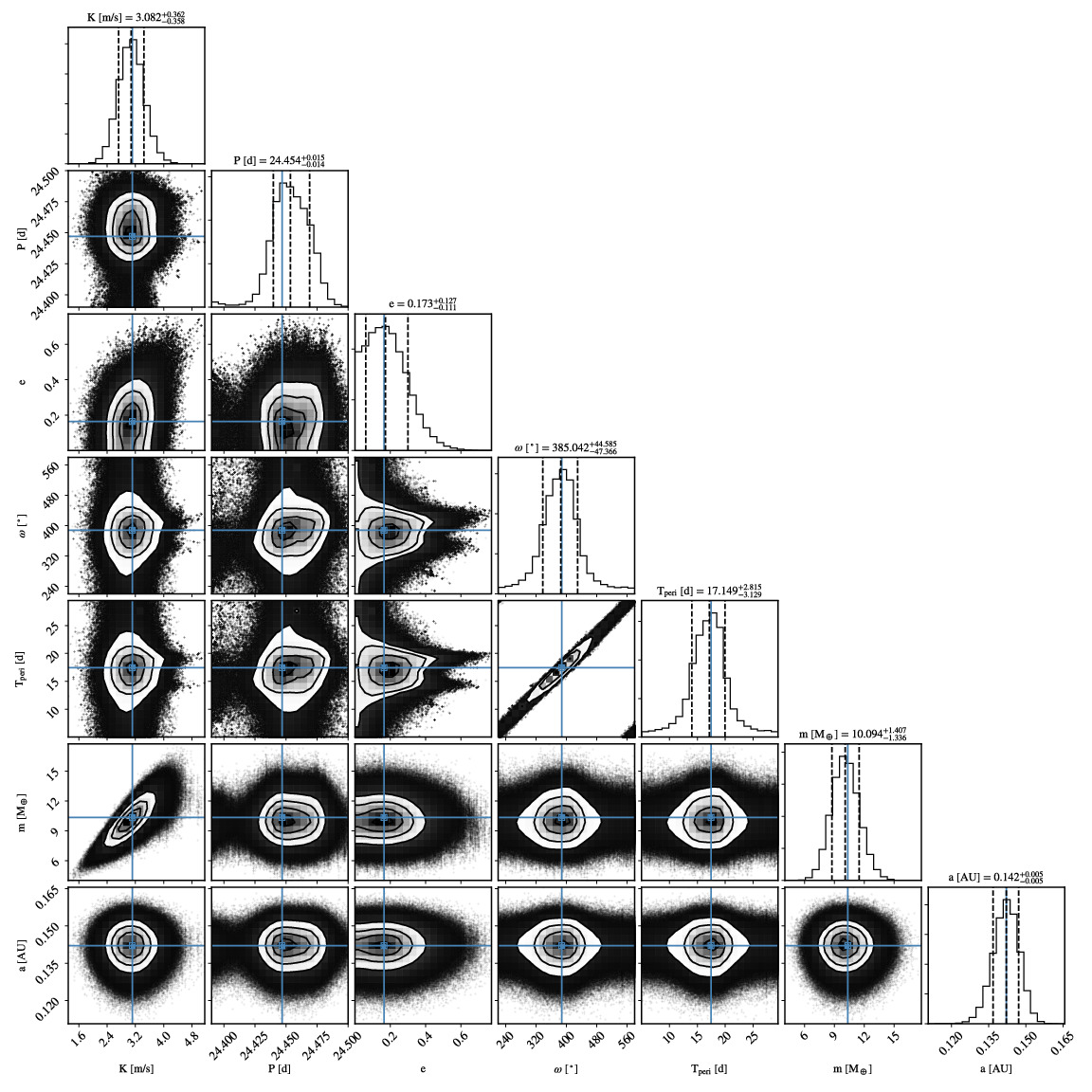}
\caption{MCMC distribution of the planet orbiting GJ\,338\,B. The vertical dashed lines indicate the 16\%, 50\%, and 84\% quantiles of the fitted parameters; this corresponds to 1\,$\sigma$ uncertainty. The best fit is colored.
}
\label{fig:corner_plot_GP-eps-converted-to.jpg}
\end{figure*}

%--------------------------------------------------------------------
%--------------------------------------------------------------------
%--------------------------------------------------------------------
%--------------------------------------------------------------------
%--------------------------------------------------------------------
%--------------------------------------------------------------------

\clearpage

\newpage
\clearpage
\section{Tables}

Tables \ref{CARMENES_rv_measurments}--\ref{tab:ref astrometric data} are only available at the CDS (see title footnote on page 1).

\longtab[1]{
\begin{landscape}
\begin{scriptsize}
\begin{longtable}{lcccccccccc}
\caption{\label{CARMENES_rv_measurments} GJ 338 B data of the CARMENES observations.}\\
\hline\hline
\noalign{\smallskip}
BJD & RV$_{\rm VIS}$ & dLW & H$\alpha$  & FWHM & Contrast & BIS & IRT1 & IRT2 & IRT3 & CRX \\
(d)  & (m\,s$^{-1}$)  &  (m$^{2}$\,s$^{-2}$)  & & (km\,s$^{-1}$) &  & (km\,s$^{-1}$)   &  & &  & (m\,s$^{-1}$\,dex$^{-1}$) \\

 \noalign{\smallskip}
\hline
\noalign{\smallskip}
\endfirsthead
\caption{continued.}\\
\hline\hline
\noalign{\smallskip}
BJD & RV$_{VIS}$ & dLW & H$\alpha$  & FWHM & Contrast & BIS & IRT1 & IRT2 & IRT3 & CRX \\
(d)  & (m\,s$^{-1}$)  &  (m$^{2}$\,s$^{-2}$)  & & (km\,s$^{-1}$) &  & (km\,s$^{-1}$)   &  & &  & (m\,s$^{-1}$\,dex$^{-1}$) \\

\noalign{\smallskip}
\hline
\endhead
\hline
\endfoot
%Added by TeX Support
  2457395.64114  &  4.62   $\pm$  1.27   & -5.67   $\pm$ 1.73   & 0.7326  $\pm$ 0.0006  & 5.58  $\pm$ 0.01      & 24.49     $\pm$ 0.04        &   0.043   $\pm$ 0.003   &  0.581  $\pm$ 0.001  &  0.467  $\pm$ 0.001  &  0.45   $\pm$ 0.001  &  -13.17   $\pm$ 9.96  \\
  2457398.53654  &  -5.57  $\pm$  2.07   & -2.36   $\pm$ 1.8    & 0.7267  $\pm$ 0.0008  & 5.59  $\pm$ 0.01      & 24.54     $\pm$ 0.04        &   0.043   $\pm$ 0.004   &  0.579  $\pm$ 0.001  &  0.468  $\pm$ 0.001  &  0.448  $\pm$ 0.001  &  -18.66   $\pm$ 13.59 \\
  2457400.60073  &  0.76   $\pm$  1.4    & 4.63    $\pm$ 1.48   & 0.7434  $\pm$ 0.0006  & 5.58  $\pm$ 0.01      & 24.38     $\pm$ 0.04        &   0.035   $\pm$ 0.003   &  0.589  $\pm$ 0.001  &  0.477  $\pm$ 0.001  &  0.456  $\pm$ 0.001  &  -2.77    $\pm$ 10.42 \\
  2457408.64378  &  -1.63  $\pm$  3.46   & 2.49    $\pm$ 2.04   & 0.7474  $\pm$ 0.001   & 5.59  $\pm$ 0.01      & 24.34     $\pm$ 0.04        &   0.052   $\pm$ 0.005   &  0.599  $\pm$ 0.001  &  0.485  $\pm$ 0.001  &  0.466  $\pm$ 0.001  &  -20.92   $\pm$ 14.25 \\
  2457408.65378  &  0.9    $\pm$  3.25   & 13.02   $\pm$ 1.7    & 0.7481  $\pm$ 0.0005  & 5.6   $\pm$ 0.01      & 24.35     $\pm$ 0.04        &   0.043   $\pm$ 0.003   &  0.597  $\pm$ 0.001  &  0.485  $\pm$ 0.001  &  0.465  $\pm$ 0.001  &  8.31     $\pm$ 7.33  \\
  2457410.52863  &  -3.46  $\pm$  1.57   & 6.78    $\pm$ 1.36   & 0.7501  $\pm$ 0.0004  & 5.61  $\pm$ 0.01      & 24.38     $\pm$ 0.04        &   0.046   $\pm$ 0.002   &  0.593  $\pm$ 0.001  &  0.481  $\pm$ 0.001  &  0.461  $\pm$ 0.001  &  9.48     $\pm$ 8.76  \\
  2457410.53606  &  -1.92  $\pm$  1.5    & 7.35    $\pm$ 1.5    & 0.7492  $\pm$ 0.0004  & 5.6   $\pm$ 0.01      & 24.34     $\pm$ 0.04        &   0.046   $\pm$ 0.002   &  0.593  $\pm$ 0.001  &  0.481  $\pm$ 0.001  &  0.461  $\pm$ 0.001   &  4.16     $\pm$ 8.07  \\
\end{longtable}
\end{scriptsize}
\end{landscape}
}

%--------------------------------------------------------------------
%--------------------------------------------------------------------
%--------------------------------------------------------------------
%--------------------------------------------------------------------
%--------------------------------------------------------------------
%--------------------------------------------------------------------

%GJ 338 A componet info
\longtab[2]{
\begin{longtable}{lcccccc}
\caption{\label{Tab:CARMENES_rv_measurements_Acomp} GJ 338 A data of the CARMENES observations.}\\
\hline\hline
\noalign{\smallskip}
BJD & RV$_{\rm VIS}$ & dLW & H$\alpha$  & IRT1 & IRT2 & IRT3 \\
(d)  & (m\,s$^{-1}$)  &  (m$^{2}$\,s$^{-2}$)  &  &  & &  \\

 \noalign{\smallskip}
\hline
\noalign{\smallskip}
\endfirsthead
\caption{continued.}\\
\hline\hline
\noalign{\smallskip}
BJD & RV$_{\rm VIS}$ & dLW & H$\alpha$  & IRT1 & IRT2 & IRT3 \\
(d)  & (m\,s$^{-1}$)  & (m$^{2}$\,s$^{-2}$) &  &  & &  \\

\noalign{\smallskip}
\hline
\endhead
\hline
\endfoot
%Added by TeX Support
  2457467.41671  &   -14.18  $\pm$ 2.44  &    -13.71  $\pm$ 4.53   &   0.7289  $\pm$ 0.0018   &  0.594  $\pm$ 0.002   &  0.476  $\pm$ 0.002   &  0.461  $\pm$ 0.002 \\
  2457499.37135  &   -17.72  $\pm$ 2.12  &    -18.41  $\pm$ 3.33   &   0.7309  $\pm$ 0.0007   &  0.587  $\pm$ 0.001   &  0.477  $\pm$ 0.001   &  0.46   $\pm$ 0.001 \\
  2457555.37464  &   -13.82  $\pm$ 2.46  &    -9.81   $\pm$ 4.14   &   0.7373  $\pm$ 0.0011   &  0.598  $\pm$ 0.001   &  0.493  $\pm$ 0.001   &  0.467  $\pm$ 0.001 \\
  2458105.57328  &   -2.32   $\pm$ 1.64  &    0.45    $\pm$ 2.19   &   0.7587  $\pm$ 0.001    &  0.618  $\pm$ 0.001   &  0.517  $\pm$ 0.001   &  0.491  $\pm$ 0.001 \\
  2458122.45286  &   -2.15   $\pm$ 1.87  &    11.09   $\pm$ 2.31   &   0.7512  $\pm$ 0.0009   &  0.621  $\pm$ 0.001   &  0.519  $\pm$ 0.001   &  0.494  $\pm$ 0.001 \\
  2458130.66117  &   -3.15   $\pm$ 2.09  &    14.76   $\pm$ 3      &   0.755   $\pm$ 0.0013   &  0.621  $\pm$ 0.001   &  0.519  $\pm$ 0.002   &  0.496  $\pm$ 0.001 \\
  2458131.49145  &   6.73    $\pm$ 4.94  &    -38.47  $\pm$ 12.38  &   0.7491  $\pm$ 0.0038   &  0.62   $\pm$ 0.004   &  0.52   $\pm$ 0.005   &  0.493  $\pm$ 0.005 \\
  2458132.39372  &   -7.3    $\pm$ 2.73  &    -8.62   $\pm$ 6.7    &   0.748   $\pm$ 0.002    &  0.617  $\pm$ 0.002   &  0.518  $\pm$ 0.002   &  0.491  $\pm$ 0.002 \\
\end{longtable}
}

%NIR radial velocities
\begin{table}[!h]
\renewcommand\thetable{B.4}
\caption{Relative radial velocity measurements of GJ\,338\,B with CARMENES in the NIR.}
\label{CARMENES_rv_measurments_nir}
\centering  

\begin{tabular}{l c}
\hline
\hline
\noalign{\smallskip}
Epoch [BJD]   & RV$_{\rm NIR}$\\
(d) 				& $\rm (m\, s^{-1})$ \\
\noalign{\smallskip}
\hline
\noalign{\smallskip}

2457695.74104  &   -3.06     $\pm$ 20.96\\
2457698.53211  &   -52.24    $\pm$ 31.33\\
2457699.58824  &   -7.92     $\pm$ 20.26\\
2457701.65194  &   -0.48     $\pm$ 8.62 \\
2457703.73873  &   -6.41     $\pm$ 7.35 \\
2457704.53247  &   8.08      $\pm$ 6.78 \\
\hline
\end{tabular}
\end{table}

\begin{table}
\caption*{Doppler measurements for GJ\,338\,B in the NIR -- continued.}
\centering

\begin{tabular}{l c}

\hline
\hline
\noalign{\smallskip}
Epoch [BJD]   & RV$_{\rm NIR}$   \\
(d) 				&  $\rm (m\, s^{-1})$\\
\noalign{\smallskip}
\hline
\noalign{\smallskip}

2458027.64945  &   0.79      $\pm$ 7.42 \\
2458028.70023  &   -7.12     $\pm$ 7.45 \\
2458060.60622  &   -20.71    $\pm$ 11.15\\
2458064.58946  &   -28.85    $\pm$ 7.74 \\
2458084.51992  &   3.03      $\pm$ 6.03 \\
2458085.61449  &   -0.01     $\pm$ 14.99\\
\hline

\end{tabular}    
\end{table}

%----------------------------------------------
%----------------------------------------------
%----------------------------------------------%----------------------------------------------
%----------------------------------------------
%---------RV in common CARMENES Va-Vb

\longtab[3]{
\begin{longtable}{lcccccc}
\caption{\label{Tab:CARMENES_rv_va-vb_in_common} Relative and differential RVs for binary GJ338AB computed with common template (GJ338B).}\\
\hline\hline
\noalign{\smallskip}
BJD & RV$_{\rm A}$ & eRV$_{\rm A}$ & RV$_{\rm B}$  & eRV$_{\rm B}$ & RV$_{\rm A}$ $-$ RV$_{\rm B}$ & e(RV$_{\rm A}$ $-$ RV$_{\rm B}$)\\
(d)   & (m\,s$^{-1}$)  &  (m\,s$^{-1}$)  &  (m\,s$^{-1}$) &  (m\,s$^{-1}$) &  (m\,s$^{-1}$) &  (m\,s$^{-1}$) \\

\noalign{\smallskip}
\hline
\noalign{\smallskip}
\endfirsthead
\caption{continued.}\\
\hline\hline
\noalign{\smallskip}
BJD & RV$_{\rm A}$ & eRV$_{\rm A}$ & RV$_{\rm B}$  & eRV$_{\rm B}$ & RV$_{\rm A}$-RV$_{\rm B}$ & e(RV$_{\rm A}$ $-$ RV$_{\rm B}$)\\
(d)   & (m\,s$^{-1}$)  &  (m\,s$^{-1}$)  &  (m\,s$^{-1}$) &  (m\,s$^{-1}$) &  (m\,s$^{-1}$) &  (m\,s$^{-1}$) \\

\noalign{\smallskip}
\hline
\endhead
\hline
\endfoot
%Added by TeX Support
  CARMENES & & & & & & \\
  \hline
  \noalign{\smallskip}
 2457467.40826 &  -1232.61  & 2.55 & 6.42   & 4.62 & -1239.02 & 5.27 \\
 2457499.35162 &  -1236.58  & 1.97 & -5.43  & 2.44 & -1231.14 & 3.13 \\
 2457555.35128 &  -1232.69  & 2.51 & -0.60  & 2.27 & -1232.08 & 3.38 \\
 2458118.80384 &  -1221.67  & 1.71 & -5.11  & 2.43 & -1216.55 & 2.97 \\
 2458125.93723 &  -1222.39  & 2.08 & 2.074  & 7.08 & -1224.46 & 7.38 \\
 2458135.3882  &  -1222.66  & 1.72 & -9.93  & 1.55 & -1212.72 & 2.31 \\
 2458160.32833 &  -1215.61  & 2.93 & -6.86  & 5.27 & -1208.74 & 6.03 \\
\end{longtable}
}

%----------------------------------------------
%----------------------------------------------
%----------------------------------------------%----------------------------------------------
%----------------------------------------------
%ASTROMETRY 

\begin{table}
\renewcommand\thetable{B.5}
\caption{Relative astrometric position of the binary companion GJ\,338\,B to GJ\,338\,A.}
\label{tab:astrometric data}
\centering  

\begin{tabular}{l c c l}
\hline
\hline
\noalign{\smallskip}
Epoch (a) & $\theta$ ($\deg$) & $\rho$ ($''$) & Ref. $^{(a)}$\\
\noalign{\smallskip}
\hline
\noalign{\smallskip}

1821.50   &  43.8   &  21.12   &  StF1837 \\ 
1824.46   &  44.2   &  20.80   &  S\_\_1906A \\ 
1831.13   &  51.5   &  20.   &  HJ\_1833c \\ 
1833.54   &  48.5   &  20.03   &  StF1837 \\ 
1840.32   &  50.6   &  20.15   &  Stt1878 \\ 
1841.44   &  50.5   &  16.68   &  Mad1842a \\ 
1841.54   &  50.6   &  20.32   &  Da\_1867 \\
\hline
\end{tabular}
\end{table}

\begin{table}
\caption*{Relative astrometric position -- continued.}
\centering
\begin{tabular}{l c c l}
\hline
\hline
\noalign{\smallskip}
Epoch (a) & $\theta$ ($\deg$) & $\rho$ ($''$) & Ref. $^{(a)}$\\
\noalign{\smallskip}
\hline
\noalign{\smallskip}

1908.32   &  66.8   &  19.26   &  Lau1908c \\ 
1908.32   &  66.9   &  19.07   &  Hzg1909 \\ 
1908.34   &  66.6   &  19.10   &  Prz1908 \\ 
1909.30   &  67.7   &  18.95   &  Dob1927 \\ 
1909.39   &  66.9   &  19.17   &  Lau1909b \\ 
1911.23   &  67.4   &  19.10   &  Lau1911b \\ 
1912.22   &  67.8   &  18.88   &  Vou1922 \\
\hline
\end{tabular}
\end{table}

\begin{table}
\caption*{Relative astrometric position -- continued.}
\centering
\begin{tabular}{l c c l}
\hline
\hline
\noalign{\smallskip}
Epoch (a) & $\theta$ ($\deg$) & $\rho$ ($''$) & Ref. $^{(a)}$\\
\noalign{\smallskip}
\hline
\noalign{\smallskip}

1939.27   &  74.9   &  18.40   &  Scd1940a \\ 
1939.62   &  75.03   &  18.477   &  Chg1972 \\ 
1940.04   &  75.34   &  18.444   &  Str1946 \\ 
1941.13   &  75.63   &  18.428   &  Str1946 \\ 
1942.27   &  76.7   &  18.43   &  Dur1944b \\ 
1943.00   &  76.3   &  18.14   &  Baz1945a \\ 
1944.16   &  76.30   &  18.400   &  Chg1972 \\
\hline
\end{tabular}
\end{table}

\begin{table}
\caption*{Relative astrometric position -- continued.}
\centering
\begin{tabular}{l c c l}
\hline
\hline
\noalign{\smallskip}
Epoch (a) & $\theta$ ($\deg$) & $\rho$ ($''$) & Ref. $^{(a)}$\\
\noalign{\smallskip}
\hline
\noalign{\smallskip}

1958.283   &  80.37   &  18.063   &  Ge\_1964 \\ 
1958.299   &  80.39   &  18.087   &  Ge\_1964 \\ 
1959.86   &  80.95   &  18.041   &  Kpr1992 \\ 
1960.67   &  81.18   &  18.050   &  Chg1972 \\ 
1961.44   &  81.23   &  18.01   &  Hau1965 \\ 
1962.173   &  81.820   &  18.022   &  PkO2003 \\ 
1962.173   &  81.82   &  18.022   &  PkO1988 \\
\hline
\end{tabular}
\end{table}

\begin{table}
\caption*{Relative astrometric position -- continued.}
\centering
\begin{tabular}{l c c l}
\hline
\hline
\noalign{\smallskip}
Epoch (a) & $\theta$ ($\deg$) & $\rho$ ($''$) & Ref. $^{(a)}$\\
\noalign{\smallskip}
\hline
\noalign{\smallskip}

1970.032   &  83.76   &  17.861   &  USN1974 \\ 
1970.032   &  83.76   &  17.864   &  USN1974 \\ 
1970.032   &  83.78   &  17.869   &  USN1974 \\ 
1970.035   &  83.81   &  17.875   &  USN1974 \\ 
1970.035   &  83.87   &  17.865   &  USN1974 \\ 
1970.082   &  83.83   &  17.852   &  USN1974 \\ 
1970.082   &  83.88   &  17.863   &  USN1974 \\
\hline
\end{tabular}
\end{table}

\begin{table}
\caption*{Relative astrometric position -- continued.}
\centering
\begin{tabular}{l c c l}
\hline
\hline
\noalign{\smallskip}
Epoch (a) & $\theta$ ($\deg$) & $\rho$ ($''$) & Ref. $^{(a)}$\\
\noalign{\smallskip}
\hline
\noalign{\smallskip}
1971.242   &  84.18   &  17.833   &  USN1974 \\ 
1971.242   &  84.19   &  17.838   &  USN1974 \\ 
1971.264   &  84.18   &  17.839   &  USN1974 \\ 
1971.264   &  84.22   &  17.837   &  USN1974 \\ 
1971.264   &  84.22   &  17.845   &  USN1974 \\ 
1971.264   &  84.24   &  17.839   &  USN1974 \\ 
1971.277   &  84.16   &  17.840   &  USN1974 \\
\hline
\end{tabular}
\end{table}

\begin{table}
\caption*{Relative astrometric position -- continued.}
\centering
\begin{tabular}{l c c l}
\hline
\hline
\noalign{\smallskip}
Epoch (a) & $\theta$ ($\deg$) & $\rho$ ($''$) & Ref. $^{(a)}$\\
\noalign{\smallskip}
\hline
\noalign{\smallskip}
1975.103   &  85.37   &  17.758   &  USN1978 \\ 
1975.149   &  85.599   &  17.780   &  PkO2003 \\ 
1975.150   &  85.60   &  17.777   &  PkO1988 \\ 
1975.151   &  85.375   &  17.765   &  PkO2016 \\ 
1975.179   &  85.381   &  17.767   &  PkO2014b \\ 
1975.182   &  85.38   &  17.768   &  USN1978 \\ 
1975.272   &  85.41   &  17.745   &  USN1978 \\
\hline
\end{tabular}
\end{table}

\begin{table}
\caption*{Relative astrometric position -- continued.}
\centering
\begin{tabular}{l c c l}
\hline
\hline
\noalign{\smallskip}
Epoch (a) & $\theta$ ($\deg$) & $\rho$ ($''$) & Ref. $^{(a)}$\\
\noalign{\smallskip}
\hline
\noalign{\smallskip}
1977.956   &  86.19   &  17.711   &  USN1984 \\ 
1977.956   &  86.20   &  17.702   &  USN1984 \\ 
1978.080   &  86.244   &  17.707   &  PkO2016 \\ 
1978.083   &  86.548   &  17.735   &  PkO2003 \\ 
1978.087   &  86.51   &  17.728   &  PkO1988 \\ 
1978.092   &  84.0   &  18.10   &  Scn2003 \\ 
1978.096   &  86.234   &  17.707   &  PkO2014b \\
\hline
\end{tabular}
\end{table}

\begin{table}
\caption*{Relative astrometric position -- continued.}
\centering
\begin{tabular}{l c c l}
\hline
\hline
\noalign{\smallskip}
Epoch (a) & $\theta$ ($\deg$) & $\rho$ ($''$) & Ref. $^{(a)}$\\
\noalign{\smallskip}
\hline
\noalign{\smallskip}
1990.045   &  89.5   &  17.10   &  Tob2003 \\ 
1990.260   &  89.939   &  17.499   &  PkO2016 \\ 
1990.262   &  89.96   &  17.495   &  PkO2010 \\ 
1990.262   &  89.919   &  17.490   &  PkO2014b \\ 
1990.33   &  89.0   &  17.69   &  Kzn1994 \\ 
1991.25   &  90.1   &  17.47   &  HIP1997a \\ 
1991.81   &  90.5   &  17.476   &  TYC2000b \\
\hline
\end{tabular}
\end{table}

\begin{table}
\caption*{Relative astrometric position -- continued.}
\centering
\begin{tabular}{l c c l}
\hline
\hline
\noalign{\smallskip}
Epoch (a) & $\theta$ ($\deg$) & $\rho$ ($''$) & Ref. $^{(a)}$\\
\noalign{\smallskip}
\hline
\noalign{\smallskip}
2012.954   &  96.4   &  17.2   &  CrC2014 \\ 
2013.392   &  97.8   &  16.99   &  Ary2014 \\ 
2014.2070   &  97.373   &  17.1103   &  Izm2015 \\ 
2014.237   &  97.5   &  17.08   &  WSI2015 \\ 
2015.278   &  97.2   &  17.22   &  Ctt2016 \\ 
2015.283   &  97.73   &  17.184   &  CDa2016 \\ 
2015.2937   &  97.80   &  17.09   &  Cve2017 \\ 
2016.213   &  98.1   &  17.102   &  WSI2017b \\ 
2016.233   &  98.19   &  17.007   &  Loc2017a \\ 
2016.369   &  97.4   &  17.2   &  Wbt2017 \\ 
2017.130   &  99.0   &  17.078   &  WSI2018a \\ 
2017.266   &  97.9   &  17.28   &  Ctt2018 \\ 
2017.268   &  97.1   &  17.1   &  Wbt2018 \\ 
2017.28   &  98.17   &  17.17   &  CC\_2018a \\ 
%NEW data
2019.142 & 101.95 & 16.99 & This work\\
2019.169 & 102.14 & 17.01 & This work\\
2019.185 & 102.14 & 16.98 & This work\\
2019.145 & 102.24 & 17.03 & This work\\
2019.251 & 101.27 & 17.02 & This work\\
2019.256 & 101.31 & 17.02 & This work\\
2019.281 & 101.40 & 17.02 & This work\\
\hline
\end{tabular}
\tablefoot{\centering $^{(a)}$ References listed in Table \ref{tab:ref astrometric data}. }
\end{table}

%--------------------------------------------------------------------
%--------------------------------------------------------------------
%--------------------------------------------------------------------
%--------------------------------------------------------------------
%--------------------------------------------------------------------
%--------------------------------------------------------------------

%REF astrometric data

\longtab[6]{
\begin{longtable}{l l}
\caption{\label{tab:ref astrometric data} References in Table \ref{tab:astrometric data}. }\\
\hline
\hline
\noalign{\smallskip}
Ref.   &   \\
\noalign{\smallskip}
\hline
\noalign{\smallskip}
\endfirsthead
\caption{Continued.} \\
\hline
\hline
\noalign{\smallskip}
Ref.   &   \\
\noalign{\smallskip}
\hline
\noalign{\smallskip}
\endhead
\hline
\endfoot
\hline
\endlastfoot

AbH1973    &    Abt, H.A. \&  Levy, S.G.                                                                    \\    
           &    AJ 78, 1093, 1973                                                                         \\    
Abt1922    &    Abetti, G.                                                                                \\    
           &    Pub. R. Obs. Astrof. Arcetri 39, 3, 1922                                                  \\    
ADP1998    &    Abad, C., Docobo, J., \&  Della Prugna, F.                                                  \\    
           &    A\& AS 133, 71, 1998                                                                        \\    
Alz2003b   &    Alzner, A.                                                                                \\    
           &    Webb Soc., Double Star Circ. 11, 2003                                                     \\
\end{longtable}
  
}% End longtab      

%----------------------------------------------

\end{appendix}
           
\end{document}